\documentclass[12pt, draftclsnofoot, onecolumn]{IEEEtran}
\usepackage{epsfig,graphicx,subfigure,psfrag,amsmath,cases,bm}
\usepackage{latexsym,amssymb,algorithm,mathtools}
\usepackage{algorithmic}
\usepackage{color}
\usepackage{url}
\usepackage{scrtime}
\usepackage{stfloats}
\usepackage{tablefootnote}
\usepackage{cite}
\usepackage{amsfonts,mathabx}
\usepackage{textcomp}
\usepackage{xcolor,capt-of}
\usepackage{multirow}
\allowdisplaybreaks

\author{\IEEEauthorblockN {D. Xu, X. Yu, D. W. K. Ng, A. Schmeink, and R. Schober}

\thanks{Dongfang Xu and Robert Schober are with the Institute for Digital Communications, Friedrich-Alexander-University Erlangen-N\"urnberg (FAU), Germany (email:\{dongfang.xu, robert.schober\}@fau.de). Xinghao Yu is with Department of Electronic and Computer Engineering, The Hong Kong University of Science and Technology, Hong Kong (email: eexyu@ust.hk). Derrick Wing Kwan Ng is with the School of Electrical Engineering and Telecommunications, the University of New South Wales, Australia (email: w.k.ng@unsw.edu.au). Anke Schmeink is with the Chair of Information Theory and Data Analytics, RWTH Aachen University, Germany (email: schmeink@inda.rwth-aachen.de).}
}

\newtheorem{T-Prob}{Transformed Problem}

\DeclareMathOperator{\maxo}{maximize}

\DeclareMathOperator{\st}{subject\hspace*{1.4mm}to}

\newtheorem{Remark}{Remark}

\let\oldbibliography\thebibliography
\renewcommand{\thebibliography}[1]{%
  \oldbibliography{#1}%
  \setlength{\itemsep}{0pt}%
}

\title{Robust and Secure Resource Allocation for ISAC Systems: A Novel Optimization Framework for Variable-Length Snapshots}

\begin{document}
\maketitle
\begin{abstract}
In this paper, we study the robust resource allocation design for secure communication in an integrated sensing and communication (ISAC) system. A multi-antenna dual-functional radar-communication (DFRC) base station (BS) serves multiple single-antenna legitimate users and senses for targets simultaneously, where already identified targets are treated as potential single-antenna eavesdroppers. The DFRC BS scans a sector with a sequence of dedicated beams, and the ISAC system takes a snapshot of the environment during the transmission of each beam. Based on the sensing information, the DFRC BS can acquire the channel state information (CSI) of the potential eavesdroppers. Different from existing works that focused on the resource allocation design for a single snapshot, in this paper, we propose a novel optimization framework that jointly optimizes the communication and sensing resources over a sequence of snapshots with adjustable durations. Besides, artificial noise (AN) is exploited by the BS for joint sensing and physical layer security provisioning. To this end, we jointly optimize the duration of each snapshot, the beamforming vector, and the covariance matrix of the AN for maximization of the system sum secrecy rate over a sequence of snapshots while guaranteeing a minimum required average achievable rate and a maximum information leakage constraint for each legitimate user. The resource allocation algorithm design is formulated as a non-convex optimization problem, where we account for the imperfect CSI of both the legitimate users and the potential eavesdroppers. To make the problem tractable, we derive a bound for the uncertainty region of the potential eavesdroppers' small-scale fading based on a safe approximation, which facilitates the development of a block coordinate descent-based iterative algorithm for obtaining an efficient suboptimal solution. Simulation results illustrate that the proposed scheme can significantly enhance the physical layer security of ISAC systems compared to three baseline schemes. Moreover, compared to the conventional multi-stage approach for ISAC system design, the proposed approach based on variable-length snapshots not only facilitates a highly-directional offline sensing beam design but also allows us to flexibly prioritize communication or sensing depending on the application scenario.
\vspace*{0mm}
\end{abstract}
\section{Introduction}
During the past decades, the radio spectrum crunch has become a severe problem for wireless communications~\cite{wong2017key,7289385}. This has spurred both academia and industry to seek new promising solutions to address this issue. Recently, there has been a significant research interest in integrated sensing and communication (ISAC), which is expected to dramatically improve spectral efficiency \cite{8999605,9540344,9755276}. In particular, supported by novel dual-functional radar-communication (DFRC) base stations (BSs), ISAC not only enables the spectral coexistence of radar systems and wireless communication systems, but also allows the dual use of expensive hardware equipment. Moreover, by embedding information into radar sensing signals or vice versa, ISAC can simultaneously provide communication services for users and acquire information of sensing targets. Motivated by these advantages, several works have proposed to integrate DFRC BSs into wireless systems and to exploit ISAC techniques to enhance communication and sensing performance at low costs  \cite{9557830,8641142,8386661,9685478,8454491,9737357,9729087}. For instance, the authors of \cite{8386661} developed optimal waveforms for DFRC systems to achieve a flexible trade-off between radar and communication performance. Also, in \cite{9685478}, the authors optimized the transmit beamforming vectors for the  minimization of the deviation of the radar sensing beam from a desired sensing beam pattern, while ensuring the quality-of-service (QoS) requirements of communication users. Although \cite{8641142,8386661,9685478,8454491,9737357,9729087} have revealed the enormous potential of ISAC to facilitate joint sensing and information transmission in a spectrum- and energy-efficient manner, they have not considered security which is of utmost importance in wireless communication systems.
\par
Due to the broadcast nature of the wireless medium, wireless communication systems are inherently vulnerable to potential security breaches. Yet, the integration of sensing and communication further aggravates the vulnerability to eavesdropping which creates a bottleneck for unlocking the full potential of ISAC systems \cite{8999605,9755276,9013227}. Specifically, for radar sensing, the DFRC BS has to illuminate the targets with high power such that the desired sensing information, e.g., angle and velocity, can be obtained based on their echoes \cite{1183861}. Also, the side lobes of the beam used for sensing should be suppressed below a desired threshold to avoid unfavorable clutter. To facilitate joint high-quality sensing and communication, the DFRC BS has to design the information-carrying signal to synthesize such a highly-directional sensing beam. As a result, the information-carrying signal is inevitably leaked to the targets and can potentially be decoded for wiretapping if the targets are malicious eavesdroppers. These observations make security a main concern for the design of practical ISAC systems. 
\par
The injection of artificial noise (AN) at the BS has been widely used to improve physical layer security in wireless communication systems \cite{negi2005secret,yu2019robust}. Specifically, by exploiting channel state information (CSI), the AN signal can be designed to intentionally degrade the channel quality of the eavesdropper without considerably affecting the signal-to-interference-plus-noise ratio (SINR) at the legitimate users. The extra degrees of freedom (DoFs) introduced by AN injection are also beneficial for security provisioning in ISAC systems. In particular, given a desired sensing area, the radar system can pre-design the beamwidth, directivity, and sidelobe level of a set of beams taking into account the geometry of the target area and the required sensing accuracy. These beams can then be sequentially transmitted to facilitate fine resolution target scanning \cite{4516997}. As a result, the DFRC BS can synthesize the desired sensing signal by optimizing the amalgamation of the information-carrying signal and the AN signal, which facilitates simultaneous sensing and secure communication \cite{9755276}. Initial results on joint beamforming and AN design for secure ISAC systems have been presented in \cite{9199556,ren2021optimal}. In particular, the authors of \cite{9199556} considered a multiple-input multiple-output DFRC system and studied the joint design of the beamforming vectors and the AN covariance matrix for minimization of the SINR at a radar target (potential eavesdropper) while meeting the QoS requirements of the legitimate users. Besides, in \cite{ren2021optimal}, the authors investigated the optimal transmit beamforming design for minimization of the mismatch between the actual beam pattern and the desired beam pattern to ensure secure communication in an ISAC system. However, the techniques developed in \cite{9199556,ren2021optimal} may not be capable of exploiting the full potential of ISAC due to the adopted multi-stage design approach. In particular, in the first stage, the DFRC BS generates an omnidirectional beam to sense the presence of targets by analyzing the echo signals \cite{8386661}. Then, based on this analysis, the DFRC BS designs a multi-beam beam pattern whose main lobes cover the directions of all targets \cite{4516997}. At last, the DFRC BS optimizes the beamforming vectors for the data signals and the AN to synthesize the desired multi-beam beam pattern for joint sensing and secure communication. Alternatively, the DFRC BS may directly exploit the rough sensing information acquired in the first stage to optimize the actual beam pattern \cite{ren2021optimal,9124713}. Yet, as this requires the DFRC BS to directly synthesize a flat-top beam pattern, unavoidable ripples in the main lobe appear which potentially degrade the sensing quality \cite{4516997}. Herein, we refer to the transmission of the actual beam pattern as one snapshot of the ISAC system. In fact, the multi-stage design approach outlined above has been adopted in most existing works, e.g., \cite{9124713,8386661,9685478,8454491,9737357,9729087}. However, this approach has several drawbacks. Specifically, as mentioned before, in practical radar systems, the ideal sensing beam pattern has to have a high directivity to illuminate the targets with high power while avoiding unfavorable clutter \cite{1399141}. Yet, the isotropic radiation pattern adopted in the first stage of the multi-stage design approach may cause heavy clutter returns, leading to very coarse sensing information regarding the targets \cite{4276989}. Moreover, the imprecise information obtained in the first stage may make the subsequent design of the multi-beam beam pattern ineffective, which potentially degrades the sensing quality. The above discussion suggests that the existing single snapshot multi-stage design approach may not lead to high-performance ISAC systems. Furthermore, most existing works, e.g., \cite{8641142,8386661,9685478,8454491,9737357,9729087}, \cite{9199556,ren2021optimal}, assumed a pure line-of-sight (LoS) channel between the DFRC BS and the target, which may be an overly-simplistic assumption for practical ISAC systems. Specifically, for terrestrial ISAC systems, the channel between the DFRC BS and the target is expected to include both an LoS and multi-path fading components \cite{9737357,4753823,8738892}, constituting a Ricean fading channel. Hence, the results obtained in \cite{8641142,8386661,9685478,8454491,9737357,9729087} for pure LoS ISAC channels may not be applicable in practice.
\par
Motivated by the above discussion, in this paper, we propose a novel design approach and optimization framework for ISAC systems, where the available resources are jointly optimized over a sequence of variable-length snapshots. In particular, a multi-antenna DFRC BS is tasked to scan a sector of a cell for sensing targets, while providing secure communication services to multiple single-antenna legitimate users, where targets detected in previous scans are treated as potential single-antenna eavesdroppers. In fact, periodic scanning is the fundamental mode of operation of most practical radar systems for realizing target sensing \cite{8365918}. By dividing a sector into several slices of equal size, the DFRC BS can provide high-quality sensing service by successively illuminating the individual slices with highly-directional beams. Then, based on the received echo signals, the DFRC BS can acquire the CSI of the potential eavesdroppers. Given the geometry of a slice of the sector, the beamwidth and directivity of the sensing beams can be designed offline. To facilitate joint sensing and secure communication, in each snapshot, the DFRC BS jointly optimizes the beamforming vectors and the covariance matrix of the AN to synthesize a pre-designed highly-directional beam pattern to illuminate one specific slice of the sector. Moreover, by varying the duration of the snapshots, we can flexibly prioritize communication or sensing based on the application scenario without significantly degrading the other functionality. On the other hand, in practice, the integration of radar and secure communication may give rise to CSI estimation errors at the DFRC BS originating from both the radar system and the communication system. To facilitate reliable communication and sensing, we investigate a robust resource allocation design for ISAC systems by taking into account the imperfect knowledge of the CSI of both the legitimate users and the potential eavesdroppers.
\par
The contributions of this paper can be summarized
as follows:
\begin{itemize}
\item We formulate the robust and secure resource allocation algorithm design for multiuser ISAC systems as a non-convex optimization problem taking into account the imperfect knowledge of the CSI of the legitimate users and potential eavesdroppers.  
\item We propose a novel design strategy and optimization framework for ISAC systems based on variable length snapshots where the DFRC BS can flexibly adjust the durations of the snapshots and sequentially serve each slice of the considered sector with a highly-directional beam to facilitate simultaneous high-quality sensing and communication. 
\item To facilitate reliable ISAC in practical systems, we first adopt a bounded uncertainty model to capture the effect of the distance estimation error. Moreover, we capture the joint impact of the angle uncertainty and the multi-path fading uncertainty of the eavesdroppers by a bounded uncertainty model, leading to an intractable uncertainty region. To overcome this difficulty, we derive a tractable bound for the joint angle and multi-path fading uncertainty region via a safe approximation and replace the original optimization problem with a lower bound. To obtain an efficient suboptimal solution, a low-complexity suboptimal algorithm is developed by capitalizing on block coordinate descent (BCD), inner approximation (IA), and semidefinite relaxation (SDR) methods.
\item 
Simulation results show that the proposed scheme can achieve a higher sum secrecy rate compared to three baseline schemes. Moreover, our results unveil the importance of taking into account CSI uncertainty for the design of ISAC systems. Besides, compared to the commonly-adopted single snapshot multi-stage design approach, e.g., \cite{9124713,8386661,9685478,8454491,9737357,9729087}, the proposed optimization framework allows us to prioritize sensing and communication in a more flexible manner.
\end{itemize}
\par
\textit{Notation:} Vectors and matrices are denoted by boldface lower case and boldface capital letters, respectively. $\mathbb{R}^{N\times M}$ and $\mathbb{C}^{N\times M}$ denote the spaces of $N\times M$ real-valued and complex-valued matrices, respectively. $\Re\left \{ \cdot \right \}$ extracts the real part of a complex number. $|\cdot|$ and $||\cdot||_2$ denote the absolute value of a complex scalar and the Euclidean norm of a vector, respectively. $\mathbf{I}_{N}$ refers to the identity matrix of dimension $N$. $\mathbb{H}^{N}$ denotes the set of complex Hermitian matrices of dimension $N$. $\mathbf{A}^T$ and $\mathbf{A}^H$ refer to the transpose and conjugate transpose of matrix $\mathbf{A}$, respectively. $\mathbf{A}\succeq\mathbf{0}$ indicates that $\mathbf{A}$ is a positive semidefinite matrix. $||\mathbf{A}||_F$, $\mathrm{Tr}(\mathbf{A})$, and $\mathrm{Rank}(\mathbf{A})$ denote the Frobenius norm, trace, and rank of matrix $\mathbf{A}$, respectively. 
$\mathcal{E}\left \{ \cdot \right \}$ represents statistical expectation. $\overset{\Delta }{=}$ and $\sim$ refer to ``defined as'' and ``distributed as'', respectively. $\mathcal{CN}(0 ,\sigma^2)$ specifies the distribution of a circularly symmetric complex Gaussian random variable with mean $0$ and variance $\sigma^2$. $\mathbf{Z}^*$ refers to the optimal value of $\mathbf{Z}$. $[z]^+$ stands for $\mathrm{max}\left \{ 0,z \right \}$. The gradient vector of function $g(\mathbf{z})$ with respect to $\mathbf{z}$ is denoted by $\nabla_{\mathbf{z}} g(\mathbf{z})$.
\section{System Model}
\label{ISAC_system_model}
In this section, we first present the system and signal models for secure multiuser ISAC downlink transmission. Then, we discuss the CSI models adopted for the legitimate users and eavesdroppers, respectively.
\begin{figure}[t]
\centering
\includegraphics[width=2.8in]{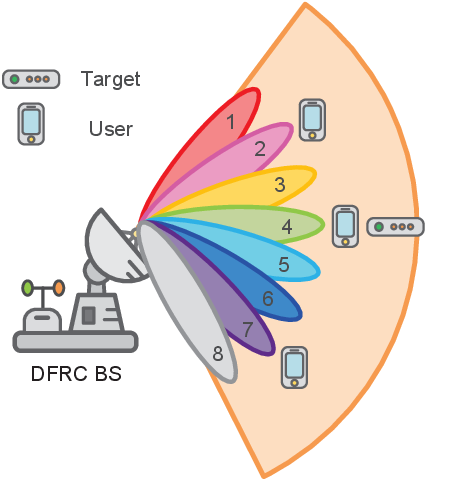}
\caption{ISAC system comprising one DFRC BS, $K=3$ legitimate users, and $J=1$ previously detected target (potential eavesdropper). To facilitate simultaneous sensing and communication, in $M=8$ snapshots, the DFRC BS sequentially transmits a set of $8$ directional beams to cover the entire sector. In each snapshot, the DFRC BS optimizes the snapshot duration, beamforming vectors, and AN such that the resulting beam facilitates joint sensing and secure communication.}
\label{system_model}
\end{figure}
\subsection{ISAC System Model}
\label{ISAC_model}
We consider a downlink ISAC multiuser communication system comprising a DFRC BS, $K$ legitimate users, and $J$ previously detected targets, cf. Figure \ref{system_model}. The DFRC BS is equipped with a uniform linear array (ULA) which comprises $N_{\mathrm{T}}$ antennas, while the $K$ legitimate users and the $J$ previously detected targets are assumed to be single-antenna devices. In particular, we consider an electronic conical scanning radar-based DFRC BS \cite{8046088,1167263}, whose task it is to scan a sector area with a sequence of $M$ beams, designed for sensing, to detect new targets and update the information already acquired about the $J$ known targets within a scanning period $T_{\mathrm{tot}}$. Different from the $K$ legitimate users that actively interact with the DFRC BS at the beginning of each scanning period, we assume that the $J$ known targets were sensed and identified by the DFRC BS in the previous scanning period, and thus, they are treated as potential eavesdroppers in the current scanning period of the ISAC system. Hence, the DFRC BS can obtain the CSI of the potential eavesdroppers based on the echoes received in the previous scanning period. Moreover, each beam is assumed to point in a specific direction to cover a slice of the sector with a desired power gain \cite{4516997}. To facilitate high-quality sensing, the DFRC BS aims at illuminating each slice of the sector with the main lobe of a highly-directional beam for a certain amount of time such that a sufficient number of echoes can be received for target information acquisition. As a result, given the geometry of the slices, we can compute the required beamwidth, directivity, and sidelobe level of the highly-directional beam in advance, and hence, the desired beam pattern can be pre-designed offline and stored at the DFRC BS. During the transmission of beam pattern $m$, the DFRC BS takes snapshot $m$ of the environment for snapshot duration $t[m]>0$, $\forall m\in\left\{ 1,\cdots,M\right\}$.\footnote{The required duration of a snapshot depends on the type of radar and the algorithm adopted for extracting the sensing information. We note that the duration of a snapshot can be varied in digitally programmable radar systems \cite{1221849}.} To facilitate joint sensing and communication, the DFRC BS generates information-carrying beamforming vectors and an AN signal to synthesize the pre-designed sensing beam patterns \cite{9199556,ren2021optimal}. Furthermore, to nimbly coordinate sensing and communication, we assume the DFRC BS can flexibly adjust the durations of the snapshots according to the requirements of the ISAC system \cite{1221849}, \cite{9393560}. For instance, if the desired sensing beam pattern in a snapshot is also favorable for secure communication, the DFRC BS can extend the duration of this snapshot to facilitate information transmission. To simplify the notation, we collect the indices of the snapshots during the entire scanning period $T_{\mathrm{tot}}$, the legitimate users, and the potential eavesdroppers in sets $\mathcal{M}=\left \{1,\ldots, M \right \}$, $\mathcal{K}=\left \{1,\ldots ,K \right \}$, and $\mathcal{J}=\left \{1,\ldots ,J \right \}$, respectively.
\subsection{Signal Model}
We focus on one scanning period of the ISAC system. In each snapshot, the DFRC BS transmits $K$ independent data streams to the $K$ legitimate users. Moreover, to facilitate target sensing while ensuring secure communication, an AN signal is added to the data signal. As a result, in the $m$-th snapshot, the signal transmitted by the DFRC BS is given by
\begin{equation}
    \mathbf{x}[m]=\underset{k\in\mathcal{K}}{\sum }\mathbf{w}_k[m]b_k[m]+\mathbf{v}[m], \forall m\in\mathcal{M}.
\end{equation}
Here, $\mathbf{w}_k[m]\in \mathbb{C}^{N_{\mathrm{T}}\times 1}$ denotes the beamformer in snapshot $m$ for serving legitimate user $k$ and $b_k[m]\in \mathbb{C}$ is the corresponding information symbol. Without loss of generality, we assume $\mathcal{E}\{\left |b_k[m] \right|^2\}=1$, $\forall\mathit{m} \in \mathcal{M}$, $\forall\mathit{k} \in \mathcal{K}$. Moreover, the AN signal $\mathbf{v}[m]\in\mathbb{C}^{N_{\mathrm{T}}\times 1}$ is modeled as
\begin{equation}
    \mathbf{v}[m]\sim \mathcal{CN}\left ( \mathbf{0}, \mathbf{V}[m] \right ),
\end{equation}
where $\mathbf{V}[m]\succeq\mathbf{0}$, $\mathbf{V}[m]\in\mathbb{H}^{N_{\mathrm{T}}}$, denotes the covariance matrix of the AN signal.
\par
The received signal at legitimate user $k$ is given by
\begin{equation}
\label{ISAC_user_signal}
    y_{\mathrm{U}_k}[m]=\underbrace{\mathbf{h}_k^H[m]\mathbf{w}_k[m]b_k[m]}_{\text{Desired signal}}+\underbrace{\underset{\substack{r\in\mathcal{K}\setminus\left\{k\right\}}}{\sum}\mathbf{h}_k^H[m]\mathbf{w}_r[m]b_r[m]}_{\text{Multiuser interference}}+\underbrace{\mathbf{h}_k^H[m]\mathbf{v}[m]}_{\text{Artificial noise}}+n_{\mathrm{U}_k}[m],
\end{equation}
where $\mathbf{h}_k[m]\in \mathbb{C}^{N_{\mathrm{T}}\times 1}$ denotes the channel vector between the DFRC BS and legitimate user $k$ in snapshot $m$, and $n_{\mathrm{U}_k}[m]\sim\mathcal{CN}(0,\sigma_{\mathrm{U}_k}^2)$ denotes the additive white Gaussian noise (AWGN) at legitimate user $k$ with variance $\sigma_{\mathrm{U}_k}^2$.
\par
The received signal at potential eavesdropper $j$ in snapshot $m$ is given by
\begin{equation}
    y_{\mathrm{E}_j}[m]=\mathbf{g}_j^H[m]\underset{k\in\mathcal{K}}{\sum}\mathbf{w}_k[m]b_k[m]+\mathbf{g}_j^H[m]\mathbf{v}[m]+n_{\mathrm{E}_j}[m],
\end{equation}
where $n_{\mathrm{E}_j}[m]\sim\mathcal{CN}(0,\sigma_{\mathrm{E}_j}^2)$ denotes the equivalent AWGN at potential eavesdropper $j$ with variance $\sigma_{\mathrm{E}_j}^2$. Moreover, $\mathbf{g}_j[m]\in \mathbb{C}^{N_{\mathrm{T}}\times 1}$ denotes the channel vector between the DFRC BS and potential eavesdropper $j$. For high-quality sensing, an LoS link between the DFRC BS and the target is needed \cite{8999605}. Therefore, in the ISAC literature, $\mathbf{g}_j[m]$ is usually modeled as an LoS channel \cite{9199556,ren2021optimal}. This assumption is valid when the target to be sensed is an aerial target or the radio propagation environment does not include scatterers. However, for terrestrial targets scattering is usually unavoidable, such that non-negligible multi-path fading may be present \cite{4753823,8738892}. To take into account this important aspect, in this paper, we model $\mathbf{g}_j[m]$ as Ricean fading with a strong LoS component and a multi-path fading component. Specifically, channel vector $\mathbf{g}_j[m]$ is modeled as
\begin{equation}
    \mathbf{g}_j[m]=\frac{\sqrt{\alpha}}{d_j[m]}\Big(\sqrt{\frac{\rho_j}{1+\rho_j}}\widehat{\mathbf{g}}_j[m]+\sqrt{\frac{1}{1+\rho_j}}\widetilde{\mathbf{g}}_j[m]\Big),
\end{equation}
where $\alpha=(\frac{\lambda_c}{4\pi})^2$ 
is a constant with $\lambda_c$ being the wavelength of the frequency of the carrier and $\rho_j\geq 0$ is the Ricean factor. Moreover, $d_j[m]\in\mathbb{R}$ denotes the distance between the DFRC BS and potential eavesdropper $j$ in snapshot $m$. Furthermore, $\widehat{\mathbf{g}}_j[m]\in \mathbb{C}^{N_{\mathrm{T}}\times 1}$ represents the deterministic LoS channel vector which is given by
\begin{equation}
    \widehat{\mathbf{g}}_j[m]=\Big[1,e^{j2\pi\omega\mathrm{sin}\theta_j[m]},\ldots,e^{j2\pi\omega(N_\mathrm{T}-1)\mathrm{sin}\theta_j[m]}\Big]^T,
\end{equation}
where $\omega$ denotes the normalized spacing between adjacent antennas and $\theta_j[m]$ is the angle specifying the direction of potential eavesdropper $j$. Besides, $\widetilde{\mathbf{g}}_j[m]\in \mathbb{C}^{N_{\mathrm{T}}\times 1}$ is the multi-path scattered channel vector denoted by $\widetilde{\mathbf{g}}_j[m]=[\widetilde{g}_{j,1}[m],\ldots,\widetilde{g}_{j,N_\mathrm{T}}[m]]^T$, where $\widetilde{g}_{j,n}[m]\in\mathbb{C}$ is the $n$-th component of $\widetilde{\mathbf{g}}_j[m]$, $\forall n\in\left\{1,\ldots,N_\mathrm{T} \right\}$.
\subsection{Channel State Information}
Next, we discuss how the CSI in the considered ISAC system is obtained. Specifically, the legitimate users are assumed to transmit pilot symbols to the DFRC BS to facilitate channel estimation. As a result, accurate instantaneous CSI of the legitimate users can be obtained at the beginning of the scanning period. Yet, as the DFRC BS sequentially scans the entire sector using $M$ snapshots, the CSI of the legitimate users may become outdated due to the movement of the legitimate users and/or the scatterers in the environment. In practice, the length of the scanning period $T_{\mathrm{tot}}$ can be tuned according to the specific application requirements. To capture this effect, we adopt a bounded uncertainty model for characterizing the CSI imperfection \cite{wang2009worst}. Specifically, the CSI of the links between the DFRC BS and legitimate user $k$ in snapshot $m$ is modeled as follows
\begin{eqnarray}
\mathbf{h}_k[m]=\overline{\mathbf{h}}_k+\mathbf{\Delta}\mathbf{h}_k[m],\hspace*{3mm}\Omega_k[m]\overset{\Delta }{=}\left\{\mathbf{\Delta}\mathbf{h}_k[m]\in \mathbb{C}^{N_{\mathrm{T}}\times 1}:\left\|\mathbf{\Delta}\mathbf{h}_k[m] \right\|_2\leq \mu_k[m]  \right\},\label{ISAC_CSI_uncertainty}
\end{eqnarray}
where $\overline{\mathbf{h}}_k\in \mathbb{C}^{N_{\mathrm{T}}\times 1}$ is the estimate of the channel of legitimate user $k$ at the beginning of the scanning period. The uncertainty caused by the outdated CSI of the channel of legitimate user $k$ in snapshot $m$ is modelled by $\mathbf{\Delta}\mathbf{h}_k[m]\in \mathbb{C}^{N_{\mathrm{T}}\times 1}$. Set $\Omega_k[m]$ collects all the possible CSI errors in the $m$-th snapshot, with their norms bounded by $\mu_k[m]$. We note that the bounded uncertainty model in \eqref{ISAC_CSI_uncertainty} is a suitable and commonly-adopted CSI uncertainty model to account for outdated CSI and estimation errors \cite{yu2019robust}.
\par
Next, we discuss the CSI of the potential eavesdroppers. In particular, for target sensing, one primary task is to acquire the distance and angle information of the targets \cite{8999605,9540344}. However, due to the hardware limitations of the DFRC BS, unavoidable estimation errors, movement of the targets, and/or finite beamwidth of the sensing beams, the DFRC BS may not be able to accurately measure the distance and angle information of the potential eavesdroppers \cite{9199556}. In this paper, we take these uncertainties into account for a robust resource allocation design. Specifically, we model the distance between the DFRC BS and eavesdropper $j$ in snapshot $m$ as $d_j[m]=\overline{d}_j+\Delta d_j[m]$, where $\overline{d}_j$ is the estimate of the distance between the DFRC BS and potential eavesdropper $j$ acquired at the beginning of the scanning period and $\Delta d_j[m]$ is the corresponding distance uncertainty in snapshot $m$. We assume that the distance uncertainty $\Delta d_j[m]$ is bounded by $D_j[m]$, i.e., $\left|\Delta d_j[m]\right|\leq D_j[m]$. Similarly, we model the angle uncertainty as $\theta_j[m]=\overline{\theta}_j+\Delta\theta_j[m]$, where $\overline{\theta}_j$ is the estimate of the angle of potential eavesdropper $j$ acquired at the beginning of the scanning period and $\Delta\theta_j[m]$ is the corresponding angle uncertainty in snapshot $m$. We assume that the angle uncertainty $\Delta\theta_j[m]$ is bounded by $\phi_j[m]$, i.e., $\left|\Delta\theta_j[m] \right|\leq \phi_j[m]$. Furthermore, potential eavesdroppers usually conceal themselves and do not proactively interact with the DFRC BS for CSI estimation. As a result, the DFRC BS may not be able to acquire information regarding the multi-path fading component of the eavesdroppers' channels. Hence, in this paper, we assume that information regarding the multi-path fading component of the channels between the DFRC BS and the potential eavesdroppers is not available. To capture this effect, we regard the multi-path fading component as a channel uncertainty that deviates from the strong LoS component and adopt a bounded uncertainty model to characterize the resulting uncertainty. In particular, we assume that the absolute values of the elements of the multi-path fading component, $\widetilde{g}_{j,n}[m]$, are bounded by $\beta_{j,n}[m]$, i.e., $\left|\widetilde{g}_{j,n}[m] \right|\leq \beta_{j,n}[m]$.\footnote{We note that although, for the Ricean fading model, the multipath component is not bounded, by carefully choosing a sufficiently large $\beta_{j,n}[m]$, the probability that the constraint is violated can be made arbitrarily small. The practicality of the adopted bounded uncertainty model to characterize Gaussian-type uncertainties, as is the case for Ricean fading, has been justified in \cite{lorenz2005robust}, \cite{pascual2005robust}.} In practice, the value of $\beta_{j,n}[m]$ depends on the specific system setup and the scattering in the environment and can be estimated based on field measurements or radio maps \cite{8648450}.
In summary, channel vector $\mathbf{g}_j[m]$ is modelled as follows
\begin{equation}
\label{eavs_channel1}
    \mathbf{g}_j[m]=\sqrt{\frac{\alpha}{(1+\rho_j)(\overline{d}_j+\Delta d_j[m])^2}}\Big[\sqrt{\rho_j}\big[1,\cdots,e^{j2\pi\omega(N_\mathrm{T}-1)\mathrm{sin}(\overline{\theta}_j+\Delta\theta_j[m])}\big]^T+\widetilde{\mathbf{g}}_j[m]\Big].
\end{equation}
We note that \eqref{eavs_channel1} contains three different uncertainty terms, i.e., $\Delta d_j[m]$, $\Delta\theta_j[m]$, and $\widetilde{\mathbf{g}}_j[m]$. The uncertainty term $\Delta d_j[m]$ can be tackled by employing a suitable mathematical transformation and the S-procedure. Yet, the uncertainty terms $\Delta\theta_j[m]$ and $\widetilde{\mathbf{g}}_j[m]$ are challenging for a robust resource allocation design. To overcome this obstacle, in Section IV, we derive a tractable bound that captures the joint effects of these two uncertainty terms by applying a safe approximation \cite{li2013safe}.

\section{Problem Formulation}
\label{ISAC_problem_formulation}
In this section, we formulate the robust resource allocation design as an optimization problem, after defining the adopted performance metrics.
\subsection{Performance Metrics}
According to \eqref{ISAC_user_signal}, the achievable rate (bits/s/Hz) of legitimate user $k$ in snapshot $m$ is given by
\begin{eqnarray}
R_k[m]=\mathrm{log}_2\Big(1+\frac{\left|\mathbf{h}_k^H[m]\mathbf{w}_k[m]\right|^2}{\underset{\substack{r\in\mathcal{K}\setminus\left\{k\right\}}}{\sum}\left|\mathbf{h}_k^H[m]\mathbf{w}_r[m]\right|^2+\mathrm{Tr}(\mathbf{h}_k[m]\mathbf{h}_k^H[m]\mathbf{V}[m])+\sigma_{\mathrm{U}_k}^2}\Big).
\end{eqnarray}
\par
To guarantee secure communications, we consider a worst-case scenario by assuming that the potential eavesdroppers are capable of eliminating all multiuser interference (MUI) before decoding the information intended for a specific legitimate user \cite{6860253}. As a result, in snapshot $m$, the capacity (bits/s/Hz) of the channel between the DFRC BS and potential eavesdropper $j$ for wiretapping the signal of legitimate user $k$ is given by
\begin{equation}
C_{k,j}[m]=\mathrm{log}_2\Big(1+\frac{\left|\mathbf{g}_j^H[m]\mathbf{w}_k[m]\right|^2}{\mathrm{Tr}(\mathbf{g}_j[m]\mathbf{g}_j^H[m]\mathbf{V}[m])+\sigma_{E_j}^2}\Big).
\end{equation}
Therefore, in the $m$-th snapshot, the maximum achievable secrecy rate between the BS and the $k$-th legitimate user is given by \cite{6781609}
\begin{equation}
    R^{\mathrm{sec}}_k[m]=\Big[R_k[m]-\underset{j\in\mathcal{J}}{\mathrm{max}}\hspace*{1mm}C_{j,k}[m]\Big]^+
\end{equation}
and the system sum secrecy rate in the $m$-th snapshot is given by $\underset{k\in\mathcal{K}}{\sum}R^{\mathrm{sec}}_k[m]$.

\subsection{Problem Formulation}
In this paper, we aim at maximizing the system sum secrecy rate over $M$ snapshots within scanning period $T_{\mathrm{tot}}$, while ensuring the average achievable rate of each legitimate user is above a pre-defined threshold, the information leakage to potential eavesdroppers is below a given threshold, and the mismatch between the synthesized actual beam and the ideal beam desired for sensing is less than a pre-defined threshold. To start with, we define beamforming matrix $\mathbf{W}_k[m]=\mathbf{w}_k[m]\mathbf{w}_k^H[m]$, $\forall m$, $\forall k$. The duration of each snapshot, the beamforming matrix, and the covariance matrix of the AN, i.e., $\left \{t[m],\mathbf{W}_k[m],\mathbf{V}[m]\right \}$, are jointly designed by solving the following optimization problem\footnote{We note that the number of data symbols transmitted per snapshot depends on the adopted standard. To comply with the Long-Term Evolution (LTE) standard \cite{ghosh2010fundamentals}, the DFRC BS has to transmit at least $84$ data symbols in each snapshot that is used for information transmission. With the proposed optimization framework, the minimum duration of each snapshot can be adjusted to accommodate the required minimum number of data symbols and to ensure that the desired beam pattern is well approximated by the covariance matrix of the transmit signal.}
\begin{eqnarray}
\label{prob1}
&&\hspace*{-12mm}\underset{\mathbf{W}_{k}[m],\mathbf{V}[m]\in\mathbb{H}^{N_{\mathrm{T}}},t[m]}{\maxo} \,\, \,\, \frac{1}{T_{\mathrm{tot}}}\underset{m\in\mathcal{M} }{\sum}t[m]\underset{k\in\mathcal{K}}{\sum}\Big[\underset{\Delta \mathbf{h}_k[m]\in\Omega _k[m]}{\mathrm{min}}R_k[m]-\underset{j\in\mathcal{J}}{\mathrm{max}}\underset{\substack{\left|\Delta\theta_j[m] \right|\leq \phi_j[m],\\\left|\widetilde{g}_{j,n}[m] \right|\leq \beta_{j,n}[m],\\\left|\Delta d_j[m]\right|\leq D_j[m]}}{\mathrm{max}}C_{j,k}[m]\Big]^+\notag\\
&&\hspace*{-4mm}\st\hspace*{10mm}
\mbox{C1:}\hspace*{1mm}\underset{k\in\mathcal{K}}{\sum }\mathrm{Tr}(\mathbf{W}_{k}[m])+\mathrm{Tr}(\mathbf{V}[m])\leq P^{\mathrm{max}},\hspace*{1mm}\forall m,\notag\\
&&\hspace*{25mm}
\mbox{C2:}\hspace*{1mm}\left\| \underset{k\in\mathcal{K} }{\sum}\mathbf{W}_{k}[m]+\mathbf{V}[m]-\mathbf{R}_{d}[m]\right\|_F^2\leq \delta_d[m],\hspace*{1mm}\forall m,
\notag\\
&&\hspace*{25mm}\mbox{C3:}\hspace*{1mm}\underset{m\in\mathcal{M}}{\sum}t[m]\leq T_{\mathrm{tot}},\hspace*{6mm}\mbox{C4:}\hspace*{1mm}\frac{1}{T_{\mathrm{tot}}}\underset{m\in\mathcal{M}}{\sum}t[m]\underset{\Delta \mathbf{h}_k[m]\in\Omega _k[m]}{\mathrm{min}}R_k[m]\geq \overline{R}_{\mathrm{req}_k},\hspace*{1mm}\forall k,
\notag\\
&&\hspace*{25mm}\mbox{C5:}\hspace*{1mm}\frac{1}{T_{\mathrm{tot}}}\underset{m\in\mathcal{M}}{\sum}t[m]\hspace*{1mm}\underset{j\in\mathcal{J}}{\mathrm{max}}\underset{\substack{\left|\Delta\theta_j[m] \right|\leq \phi_j[m],\\\left|\widetilde{g}_{j,n}[m] \right|\leq \beta_{j,n}[m],\\\left|\Delta d_j[m]\right|\leq D_j[m]}}{\mathrm{max}}C_{j,k}[m]\leq \overline{R}_{\mathrm{tol}_k},\hspace*{1mm}\forall k,
\notag\\
&&\hspace*{25mm}\mbox{C6:}\hspace*{1mm}
t_{\mathrm{min}}\leq t[m]\leq t_{\mathrm{max}},\hspace*{1mm}\forall m,\hspace*{4mm}\mbox{C7:}\hspace*{1mm}\mathbf{V}[m]\succeq\mathbf{0},\hspace*{1mm}\forall m,\notag
\\
&&\hspace*{25mm}\mbox{C8:}\hspace*{1mm}\mathbf{W}_k[m]\succeq\mathbf{0},\hspace*{1mm}\forall m,\hspace*{1mm}\forall k,
\hspace*{8.5mm}\mbox{C9:}\hspace*{1mm}\mathrm{Rank}(\mathbf{W}_k[m])\leq 1,\hspace*{1mm}\forall m,\hspace*{1mm}\forall k.
\end{eqnarray}
Here, the non-negative constant $P^{\mathrm{max}}$ in constraint C1 denotes the maximum transmit power of the DFRC BS in each snapshot. Moreover, to provide high-quality sensing service in snapshot $m$, the DFRC BS illuminates the targets with pre-designed highly-directional sensing beam patterns, which are characterized by the covariance matrix of the desired waveform \cite{8288677,9199556}, i.e., $\mathbf{R}_d[m]$. To facilitate simultaneous high-quality sensing and secure communication, the DFRC BS has to synthesize the desired sensing beam patterns by designing the information signal and the AN appropriately. As such, similar to \cite{8999605,9199556}, in this paper, we introduce constraint C2 which ensures that in each snapshot $m$, the difference between the covariance matrix of the desired highly-directional radar beam pattern and that of the actual signal transmitted by the DFRC BS is smaller than a pre-defined threshold $\delta_d[m]$.\footnote{Note that a smaller value of $\delta_d[m]$ reduces the maximum deviation allowed from the desired beam pattern, which is beneficial for target sensing. On the contrary, a larger value of $\delta_d[m]$ gives the DFRC BS more flexibility in designing the information beamforming vectors, potentially facilitating higher secrecy rates. As a result, $\delta_d[m]$ can be used to prioritize communication and sensing depending on the application scenario.} Constraint C3 guarantees that the duration of the $M$ snapshots does not exceed the maximum allowable scanning period. To ensure that each legitimate user enjoys satisfactory communication service within the scanning period, the minimum average achievable rate of user $k$ has to be greater than a desired rate $\overline{R}_{\mathrm{req}_k}$, as specified in constraint C4 \cite{8288677}. $\overline{R}_{\mathrm{tol}_k}$ in constraint C5 limits the maximum information leakage of each user to ensure secure communications.\footnote{We note that rather than restricting the secrecy rate of each user, the considered sum secrecy rate maximization formulation with constraints on the achievable rate and the maximum information leakage of each user provides a higher flexibility for secure resource allocation with respect to the rate requirements of different applications, e.g., multimedia services, Internet-of-Things, etc. In fact, with the proposed algorithm, we can ensure secure communication for each legitimate user.} Furthermore, constraint C6 specifies the permissible range of the duration of each snapshot. On the one hand, the minimum snapshot length, i.e., $t_{\mathrm{min}}$, accounts for two important aspects of sensing. First, due to the limitations of the hardware and the proportions of the radar, a minimum time interval between two consecutive radar snapshots is required \cite{skolnik1980introduction}. Second, to facilitate high-quality sensing, the DFRC BS has to illuminate each slice of the sector and receives echoes for target information acquisition for a certain minimum amount of time \cite{9755276}. On the other hand, constraining the maximum duration of a snapshot facilitates timely CSI updating and target sensing. $\mathbf{V}[m]\in\mathbb{H}^{N_{\mathrm{T}}}$ and constraint C7 ensure that $\mathbf{V}[m]$ is a covariance matrix. $\mathbf{W}_{k}[m]\in\mathbb{H}^{N_{\mathrm{T}}}$, $\forall m\in\mathcal{M}$, $\forall k\in\mathcal{K}$, and the constraints in C8 and C9 are imposed to ensure that $\mathbf{W}_k[m]=\mathbf{w}_k[m]\mathbf{w}_k^H[m]$ still holds after optimization of beamforming matrix $\mathbf{W}_k[m]$.
\par
\begin{Remark}
\label{ISAC_remark1}
We note that if constraint C4 is absent from \eqref{prob1}, the proposed robust resource allocation optimization problem ensures a non-negative maximum achievable secrecy rate for legitimate user $k$ in each snapshot, i.e., $R^{\mathrm{sec}}_k[m]=\Big[\underset{\Delta \mathbf{h}_k[m]\in\Omega _k[m]}{\mathrm{min}}R_k[m]-\underset{j\in\mathcal{J}}{\mathrm{max}}\underset{\substack{\left|\Delta\theta_j[m] \right|\leq \phi_j[m],\\\left|\widetilde{g}_{j,n}[m] \right|\leq \beta_{j,n}[m],\\\left|\Delta d_j[m]\right|\leq D_j[m]}}{\mathrm{max}}C_{j,k}[m]\Big]^+$, $\forall m$, even in the presence of CSI uncertainty. This is due to the fact that if the maximum achievable secrecy rate of legitimate user $k$ became negative, the DFRC BS would shut down the information transmission to user $k$ and reallocate the available power to other users. However, since constraint C4 imposes a minimum average achievable rate for each legitimate user, it is possible that the DFRC BS transmits to user $k$ even if $R^{\mathrm{sec}}_k[m]<0$ in snapshot $m$ as long as constraint C5 is satisfied. We note that this is also inevitable in conventional wireless communication systems. Nevertheless, we will show in Section V that by employing the proposed algorithm and carefully choosing the values of $\overline{R}_{\mathrm{req}_k}$ and $\overline{R}_{\mathrm{tol}_k}$, negative $R^{\mathrm{sec}}_k[m]$ do not occur in the entire range of the considered $\overline{R}_{\mathrm{req}_k}$. Therefore, in this paper, we remove the operation $[\cdot]^+$ from the objective function in \eqref{prob1}.
\end{Remark}
\par
We note that the optimization problem in \eqref{prob1} is non-convex. In particular, the non-convexity stems from the highly coupled optimization variables and the fractional SINR expression in both the objective function and constraints C4 and C5. Moreover, due to the continuous CSI uncertainty sets in the objective function and constraints C4 and C5, the considered optimization problem is a semi-infinite programming problem which involves an infinite number of constraints\footnote{To illustrate the semi-infinite programming nature of the considered problem, we take constraint C4 as an example and rewrite it equivalently as $\mbox{C4:}\hspace*{1mm}\frac{1}{T_{\mathrm{tot}}}\underset{m\in\mathcal{M}}{\sum}t[m]R_k[m]\geq \overline{R}_{\mathrm{req}_k},\hspace*{1mm}\forall k,\hspace*{1mm}\Delta \mathbf{h}_k[m]\in\Omega _k[m]$, leading to infinitely many constraints because of the continuous set $\Omega _k[m]$.} and is in general intractable for a robust resource allocation algorithm design. Besides, rank constraint C9 is also an obstacle for efficiently solving the formulated problem. There is no known systematic approach for solving \eqref{prob1} optimally in polynomial time. Therefore, in the next section, we first propose a series of transformations, and then, develop a computational-efficient BCD-based iterative algorithm to tackle the considered optimization problem.
\section{Solution of The Optimization Problem}
\label{ISAC_solution}
In this section, we tackle the optimization problem in \eqref{prob1}. In particular, based on a safe approximation, we first derive a bound for the intractable uncertainty region of the eavesdroppers' small-scale fading induced by uncertainty terms $\Delta\theta_j[m]$ and $\widetilde{\mathbf{g}}_j[m]$, and study the accuracy of the bound. Then, we transform the resulting problem into an equivalent form. Subsequently, by capitalizing on BCD theory, we divide the optimization variables into two blocks to overcome their coupling. Finally, by applying IA \cite{marks1978general} and SDR, we solve the BCD subproblems in an alternating manner.
\subsection{Bound for Uncertainty Region of Eavesdroppers' Small-Scale Fading}
\label{ISAC_safe_approximation}
\begin{figure}[t]
\begin{minipage}[b]{0.47\linewidth}
\hspace*{-5mm}\includegraphics[width=3.2in]{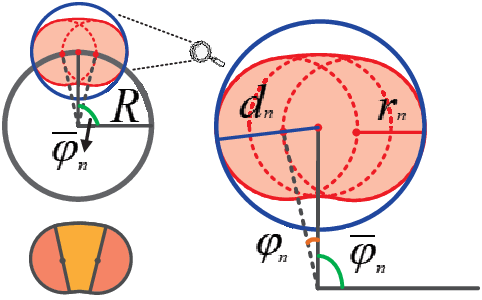}
\caption{Top-left figure: Illustration of the actual uncertainty region (the shadowed area) and the proposed safe approximation (the blue-colored circle area). Bottom-left figure: Illustration of the actual uncertainty region. Right-half figure: A zoom-in illustration of the two uncertainty regions and some key notations adopted in the geometrical analysis.}
\label{safe_appro}
\end{minipage}\hspace*{8mm}
\begin{minipage}[b]{0.47\linewidth}
\hspace*{-5mm}\includegraphics[width=3.2in]{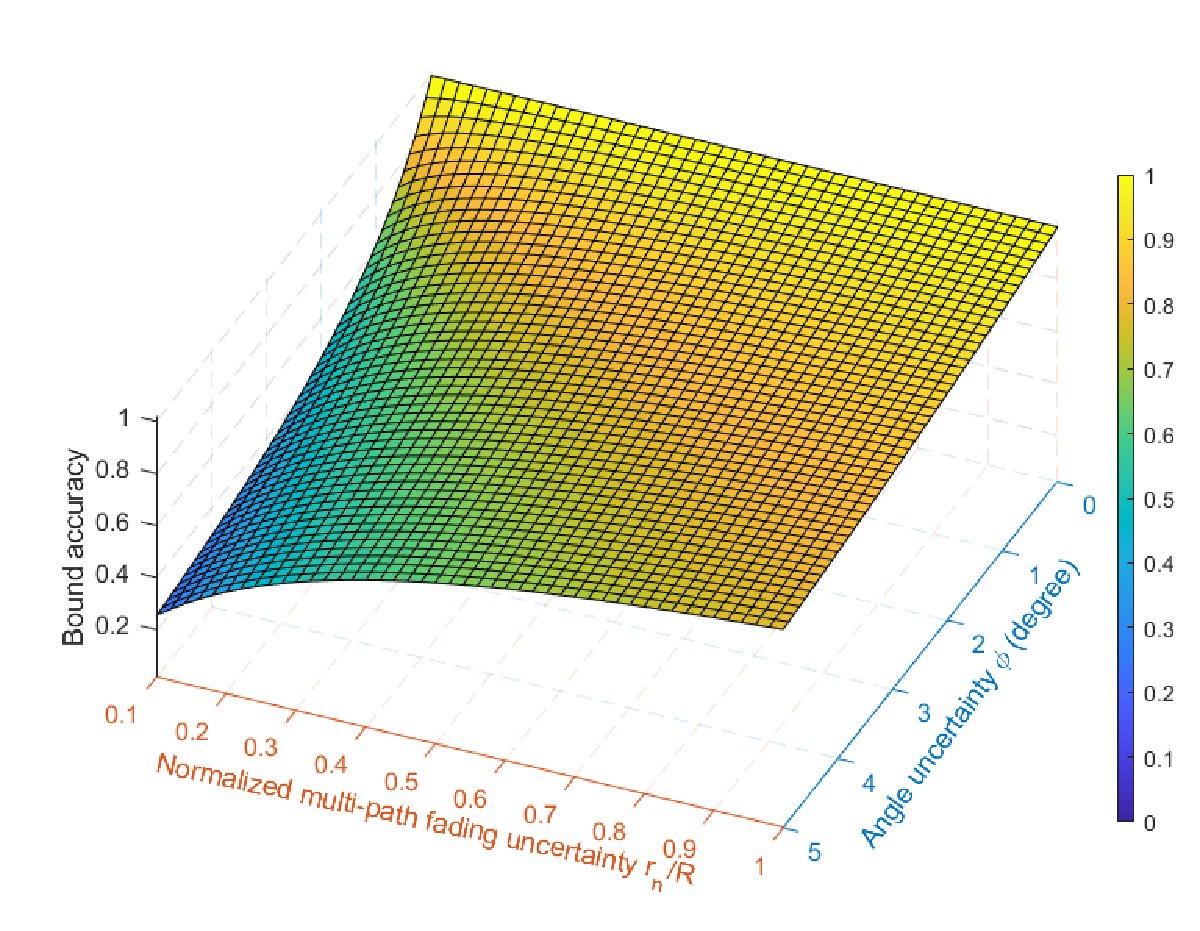}
\caption{Illustration of the tightness of the bound with respect to the angle uncertainty $\phi$ in degree and the normalized multi-path fading uncertainty $\frac{r_n}{R}$ for $n=6$, $\omega=0.5$, and $\overline{\theta}=\frac{\pi}{3}$.}
\label{ratio}
\end{minipage}
\end{figure}
In this subsection, to facilitate the robust resource allocation design, we derive a bound for the intractable uncertainty region of the eavesdroppers' small-scale fading based on analytic geometry theory.\footnote{Unless specified otherwise, in this subsection, we omit the snapshot and eavesdropper indices, i.e., $m$ and $j$, for brevity.} We start by considering the small-scale fading component of one entry of an eavesdropper's channel vector $\mathbf{g}$, i.e., $\big(\sqrt{\rho}e^{j2\pi\omega(n-1)\mathrm{sin}(\overline{\theta}+\Delta\theta)}+\widetilde{g}_n\big)$, to obtain the proposed bound with the help of Figure \ref{safe_appro}. In particular, for a given $\overline{\theta}$, we represent the term $\sqrt{\rho}e^{j2\pi\omega(n-1)\mathrm{sin}\overline{\theta}}$ as a point with angle $\overline{\varphi}_n=2\pi\omega(n-1)\mathrm{sin}\overline{\theta}$ on the black-colored circle with radius $R=\sqrt{\rho}$. Given the bounded angle uncertainty $\left|\Delta\theta\right|\leq \phi$, the term $\sqrt{\rho}e^{j2\pi\omega(n-1)\mathrm{sin}(\overline{\theta}+\Delta\theta)}$ corresponds to an arc on the black-colored circle. Considering the bounded multi-path fading uncertainty $\widetilde{g}_n$ with $\left| \widetilde{g}_n\right|\leq \beta_n$, for a given angle uncertainty $\Delta\theta$, the uncertainty region of the multi-path fading is also a circle, which has its center on the black-colored circle and radius $r_n=\beta_n$. As a result, the joint angle and multi-path fading uncertainty region is the union of a set of circles whose centers are located on the same black-colored circle. The area of this uncertainty region, denoted as $A_{\mathrm{act}_n}$, is the summation of two half circles and a part of a ring, cf. Figure \ref{safe_appro} (bottom-left). In particular, by applying analytic geometry theory, we have
\begin{equation}
    A_{\mathrm{act}_n}=2\times \frac{1}{2}\pi r^2_n+[\pi(R+r_n)^2-\pi(R-r_n)^2] \frac{2\varphi_n}{2\pi}=\pi r^2_n+4Rr_n\frac{\varphi_n}{\pi},
\end{equation}
where $\varphi_n\geq 0$ is given by
\begin{equation}
    \varphi_n=\left|2\pi\omega(n-1)\mathrm{sin}\overline{\theta}- 2\pi\omega(n-1)\mathrm{sin}(\overline{\theta}+\phi)\right|.
\end{equation}
The non-convexity of this joint uncertainty region is an obstacle to robust resource allocation design since the commonly adopted approaches for handling uncertainty such as the S-procedure \cite{boyd2004convex} require a convex uncertainty region. To overcome this difficulty, we bound the joint uncertainty region by a superset uncertainty region given by a circle, namely, the blue-colored circle in Figure \ref{safe_appro}. We note that the convex hull of the joint uncertainty region is a tighter bound than the proposed blue-colored circle. Yet, since the convex hull is difficult to model and may not be conducive to the application of the S-procedure, we adopt the more tractable circle bound in this paper. In particular, the center of the blue-colored circle is located on the black-colored circle and corresponds to $\Delta\theta=0$. By applying the law of cosines, the radius of the blue-colored circle, $d_n$, and the area of the circle
uncertainty region, $A_{\mathrm{app}_n}$, are given by, respectively,
\begin{equation}
    d_n=r_n+\sqrt{2}R\sqrt{1-\mathrm{cos}\varphi_n},\hspace*{4mm}A_{\mathrm{app}_n}=\pi(r_n+\sqrt{2}R\sqrt{1-\mathrm{cos}\varphi_n})^2.\label{ISAC_bound}
\end{equation}
Then, the accuracy of the bound can be evaluated via the ratio of the actual area and the corresponding bounding area. In particular, we have
\begin{equation}
\label{appro_ratio}
    \mathrm{ratio}(\varphi_n,r_n)=\frac{A_{\mathrm{act}_n}}{A_{\mathrm{app}_n}}=\frac{\pi r^2_n+4Rr_n\frac{\varphi_n}{\pi}}{\pi(r_n+\sqrt{2}R\sqrt{1-\mathrm{cos}\varphi_n})^2}.
\end{equation}
In Figure \ref{ratio}, we illustrate the accuracy of the bound as a function of angle uncertainty $\phi$ and normalized fading uncertainty $\frac{r_n}{R}$ for $n=6$, $\omega=0.5$, and $\overline{\theta}=\frac{\pi}{3}$. We observe that the bound is relatively accurate, especially when the angle uncertainty is small, i.e., for the high-quality sensing that is achieved by employing highly-directional beams at the DFRC BS. Note that the proposed framework not only enables the offline pre-design of highly-directional sensing beam patterns but also allows us to tune the length of the scanning period to frequently update the sensing information of the targets, which can potentially help mitigate the angle uncertainty of the eavesdroppers. As a result, the bound derived in \eqref{ISAC_bound} is particularly well-suited for the optimization framework developed in this paper.
\par
Now, $\mathbf{g}_j[m]$ in \eqref{eavs_channel1} can be modeled as the summation of a known component and a bounded uncertainty component as follows
\begin{eqnarray}
     \mathbf{g}_j[m]&\hspace*{-2mm}=\hspace*{-2mm}&\sqrt{\frac{\alpha}{(1+\rho_j)d_j^2[m]}}\Big[\sqrt{\rho_j}+\Delta g_{j,1}[m],\cdots,\sqrt{\rho_j}e^{j2\pi\omega(N_{\mathrm{T}}-1)\mathrm{sin}\overline{\theta}_j}+\Delta g_{j,N_{\mathrm{T}}}[m]\Big]^T,
\end{eqnarray}
where the uncertainty term $\Delta g_{j,n}[m]$ captures the joint effect of the angle and multi-path fading uncertainties. By replacing $r_n$, $R$, and $\mathrm{cos}\varphi_n$ in \eqref{appro_ratio} with $\beta_{j,n}[m]$, $\sqrt{\rho_j}$, and $\mathrm{cos}\big(\varphi_{j,n}[m]\big)$, respectively, we obtain a bound for the uncertainty term $\Delta g_{j,n}[m]$ as follows
\begin{equation}
    \left|\Delta g_{j,n}[m] \right|\leq \beta_{j,n}[m]+\sqrt{2\rho_j}\sqrt{1-\mathrm{cos}\big(\varphi_{j,n}[m]\big)},\label{elementwised_uncer}
\end{equation}
where $\varphi_{j,n}[m]\geq 0$ is given by
\begin{equation}
    \varphi_{j,n}[m]=\left|2\pi\omega(n-1)\mathrm{sin}\overline{\theta}_j- 2\pi\omega(n-1)\mathrm{sin}\big(\overline{\theta}_j+\phi[m]\big)\right|.
\end{equation}
Next, we collect all components $\Delta g_{j,n}[m]$, $\forall n\in\left\{1,\cdots,N_\mathrm{T} \right\}$, in vector $\mathbf{\Delta g}_j[m]\in \mathbb{C}^{N_{\mathrm{T}}\times 1}$. Since we aim at developing a robust resource allocation algorithm that jointly optimizes all elements of beamforming vector $\mathbf{w}_k[m]$, having an uncertainty region for $\mathbf{\Delta g}_j[m]$, rather than the set of uncertainty regions for the elements of $\Delta g_{j,n}[m]$ in  \eqref{elementwised_uncer}, is desirable \cite{wang2009worst,6781609}. As a result, we further propose an upper bound for the CSI uncertainty of the $j$-th potential eavesdropper $\mathbf{\Delta g}_j[m]$ as follows
\begin{eqnarray}
\label{upperbound-eaves_channel1}
    \mathbf{\Delta g}_j^H[m]\mathbf{\Delta g}_j[m] &\hspace*{-2mm}\leq\hspace*{-2mm}& \sum_{n=1}^{N_\mathrm{T}}\Big(\beta_{j,n}[m]+\sqrt{2\rho_j}\sqrt{1-\mathrm{cos}\varphi_{j,n}[m]}\Big)^2\overset{\Delta }{=}\nu_j^2[m],\\
    \Phi_j[m]&\hspace*{-2mm}\overset{\Delta }{=}\hspace*{-2mm}&\left\{\mathbf{\Delta}\mathbf{g}_j[m]\in \mathbb{C}^{N_{\mathrm{T}}\times 1}:\left\|\mathbf{\Delta}\mathbf{g}_j[m] \right\|_2\leq \nu_j[m]  \right\}.\label{upperbound-eaves_channel2}
\end{eqnarray}
Here, continuous set $\Phi_j[m]$ contains all possible small-scale fading uncertainties with their norms bounded by $\nu_j[m]>0$. Finally, we represent the channel vector of eavesdropper $j$ in snapshot $m$ as follows
\begin{equation}
    \mathbf{g}_j[m]=\sqrt{\frac{\alpha}{(1+\rho_j)d_j^2[m]}}\Big(\overline{\mathbf{g}}_j+\mathbf{\Delta g}_j[m]\Big),
\end{equation}
where $\overline{\mathbf{g}}_j\in \mathbb{C}^{N_{\mathrm{T}}\times 1}$ is the estimate of the channel of potential eavesdropper $j$ at the beginning of the scanning period and is given by
\begin{equation}
\label{overline_eaves_channel}
    \overline{\mathbf{g}}_j\overset{\Delta }{=}\Big[\sqrt{\rho_j},\sqrt{\rho_j}e^{j2\pi\omega\mathrm{sin}\overline{\theta}_j},\cdots,\sqrt{\rho_j}e^{j2\pi\omega(N_\mathrm{T}-1)\mathrm{sin}\overline{\theta}_j}\Big]^T.
\end{equation}

\subsection{Problem Reformulation}
In this subsection, we reformulate the considered optimization problem to facilitate the application of BCD. In particular, we replace the original problem \eqref{prob1} by a lower bound version which is based on the new uncertainty region in \eqref{upperbound-eaves_channel2} and given by
\begin{eqnarray}
\label{prob2}
&&\hspace*{-12mm}\underset{t[m],\mathbf{W}_{k}[m],\mathbf{V}[m]\in\mathbb{H}^{N_{\mathrm{T}}}}{\maxo} \,\, \,\, \frac{1}{T_{\mathrm{tot}}}\underset{m\in\mathcal{M} }{\sum}t[m]\underset{k\in\mathcal{K}}{\sum}\Big(\underset{\Delta \mathbf{h}_k[m]\in\Omega _k[m]}{\mathrm{min}}R_k[m]-\underset{j\in\mathcal{J}}{\mathrm{max}}\underset{\substack{\mathbf{\Delta g}_j[m]\in\Phi _j[m],\\\left|\Delta d_j[m]\right|\leq D_j[m]}}{\mathrm{max}}C_{j,k}[m]\Big)\notag\\
&&\hspace*{-4mm}\st\hspace*{10mm}
\mbox{C1-C9},
\end{eqnarray}
\par
Next, to make the objective function tractable, we first define a slack variable $\xi _k[m]\in\mathbb{R}$ to address the first semi-infinite programming term $\underset{\Delta \mathbf{h}_k[m]\in\Omega _k[m]}{\mathrm{min}}R_k[m]$ in the objective function of \eqref{prob2}, where $\xi _k[m]$ satisfies the following constraint
\begin{equation}
\mbox{C10:}\hspace*{1mm}\xi _k[m]\leq \underset{\Delta \mathbf{h}_k[m]\in\Omega _k[m]}{\mathrm{min}}R_k[m],\hspace*{1mm} \forall m,\hspace*{1mm} \forall k.
\end{equation}
To further facilitate the reformulation of \eqref{prob2}, we define another slack variable $\lambda_k[m]\in\mathbb{R}$ and rewrite constraint C10 in an equivalent manner as follows
\begin{eqnarray}
&&\hspace*{-12mm}\mbox{C10a:}\hspace*{1mm}2^{\xi _k[m]}-1\leq \lambda_k[m],\hspace*{1mm}\forall m,\hspace*{1mm} \forall k,\\
&&\hspace*{-12mm}\mbox{C10b:}\hspace*{1mm}\lambda_k[m]\leq\notag\\
&&\hspace*{-6mm}\underset{\Delta \mathbf{h}_k[m]\in\Omega _k[m]}{\mathrm{min}}\frac{\mathrm{Tr}(\mathbf{h}_k[m]\mathbf{h}_k^H[m]\mathbf{W}_k[m])}{\underset{\substack{r\in\mathcal{K}\setminus\left\{k\right\}}}{\sum}\mathrm{Tr}(\mathbf{h}_k[m]\mathbf{h}_k^H[m]\mathbf{W}_k[m])+\mathrm{Tr}(\mathbf{h}_k[m]\mathbf{h}_k^H[m]\mathbf{V}[m])+\sigma_{\mathrm{U}_k}^2 },\hspace*{1mm}\forall m,\hspace*{1mm}\forall k.
\end{eqnarray}
To handle the second semi-infinite programming term $\underset{j\in\mathcal{J}}{\mathrm{max}}\underset{\substack{\mathbf{\Delta g}_j[m]\in\Phi _j[m],\\\left|\Delta d_j[m]\right|\leq D_j[m]}}{\mathrm{max}}C_{j,k}[m]$ in the objective function, we first define two slack variables $\eta_k[m]$, $\kappa_k[m]\in\mathbb{R}$ and rewrite it as follows
\begin{eqnarray}
\label{ISAC_uncertainty2}
    &&\hspace*{-14mm}\mbox{C11:}\hspace*{1mm}\eta_k[m]\geq \mathrm{log}_2(1+\kappa_k[m]),\hspace*{1mm}\forall m,\hspace*{1mm} \forall k,\\
    &&\hspace*{-14mm}\mbox{C12:}\kappa_k[m]\geq\notag\\
    &&\hspace*{-14mm}\underset{\substack{\mathbf{\Delta g}_j[m]\in\Phi _j[m],\\\left|\Delta d_j[m]\right|\leq D_j[m]}}{\mathrm{max}}\frac{\frac{\alpha}{(1+\rho_j)(\overline{d}_j+\Delta d_j[m])^2}\mathrm{Tr}\Big(\big(\overline{\mathbf{g}}_j+\mathbf{\Delta g}_j[m]\big)\big(\overline{\mathbf{g}}_j+\mathbf{\Delta g}_j[m]\big)^H\mathbf{W}_k[m]\Big)}{\frac{\alpha}{(1+\rho_j)(\overline{d}_j+\Delta d_j[m] )^2}\mathrm{Tr}\Big(\big(\overline{\mathbf{g}}_j+\mathbf{\Delta g}_j[m]\big)\big(\overline{\mathbf{g}}_j+\mathbf{\Delta g}_j[m]\big)^H\mathbf{V}[m]\Big)+\sigma_{E_j}^2 },\forall m,j,k.
    \label{ISAC_uncertainty3}
\end{eqnarray}
Then, to decouple the uncertainty variables, we multiply simultaneously the numerator and the denominator of the SINR expression in \eqref{ISAC_uncertainty3} with $\alpha^{-1}(1+\rho_j)(\overline{d}_j+\Delta d_j[m])^2$. Subsequently, we further introduce an additional slack variable $\zeta_{k,j}[m]\in\mathbb{R}$ and reformulate constraint C12 equivalently as follows
\begin{eqnarray}
&&\hspace*{-12mm}\mbox{C12a:}\hspace*{1mm}\kappa_k[m]\mathrm{Tr}\Big(\big(\overline{\mathbf{g}}_j+\mathbf{\Delta g}_j[m]\big)\big(\overline{\mathbf{g}}_j+\mathbf{\Delta g}_j[m]\big)^H\mathbf{V}[m]\Big)\notag\\
&&\hspace*{-12mm}-\mathrm{Tr}\Big(\big(\overline{\mathbf{g}}_j+\mathbf{\Delta g}_j[m]\big)\big(\overline{\mathbf{g}}_j+\mathbf{\Delta g}_j[m]\big)^H\mathbf{W}_k[m]\Big)\geq\zeta_{k,j}[m], \hspace*{1mm}\mathbf{\Delta g}_j[m]\in\Phi _j[m],\forall m,j,k,
\\
&&\hspace*{-12mm}\mbox{C12b:}\hspace*{1mm}\zeta_{k,j}[m]\geq-\kappa_k[m]\sigma_{E_j}^2\alpha^{-1}(1+\rho_j)(\overline{d}_j+\Delta d_j[m])^2,\left|\Delta d_j[m]\right|_2\leq D_j[m],\forall m,j,k.
\end{eqnarray}
\par
As such, the optimization problem in \eqref{prob2} can be equivalently recast in hypograph form follows
\begin{eqnarray}
\label{prob3}
&&\hspace*{-12mm}\underset{\substack{t[m],\mathbf{W}_{k}[m],\mathbf{V}[m]\in\mathbb{H}^{N_{\mathrm{T}}},\\\xi_k[m],\lambda_k[m],\eta_k[m],\kappa_k[m],\zeta_{k,j}[m]}}{\maxo} \,\, \,\, \frac{1}{T_{\mathrm{tot}}}\underset{m\in\mathcal{M} }{\sum}t[m]\underset{k\in\mathcal{K}}{\sum}(\xi_k[m]-\eta_k[m])\notag\\
&&\hspace*{2mm}\st\hspace*{10mm}
\overline{\mbox{C4}}\mbox{:}\hspace*{1mm}\frac{1}{T_{\mathrm{tot}}}\underset{m\in\mathcal{M}}{\sum}t[m]\xi_k[m]\geq \overline{R}_{\mathrm{req}_k},\hspace*{1mm}\forall k,\notag\\
&&\hspace*{31mm}\overline{\mbox{C5}}\mbox{:}\hspace*{1mm}\frac{1}{T_{\mathrm{tot}}}\underset{m\in\mathcal{M}}{\sum}t[m]\eta_k[m]\leq \overline{R}_{\mathrm{tol}_k},\hspace*{1mm}\forall k,\notag\\
&&\hspace*{31mm}\mbox{C1-C3},\mbox{C6-C9},\mbox{C10a},\mbox{C10b},\mbox{C11},\mbox{C12a},\mbox{C12b}.
\end{eqnarray}
We note that by applying the aforementioned series of reformulations, both semi-infinite programming terms in the objective function of \eqref{prob2} vanish, such that the objective function becomes more manageable for a robust resource allocation algorithm design. Also, the semi-infinite constraints in C4 and C5 have been replaced by bilinear constraints $\overline{\mbox{C4}}$ and $\overline{\mbox{C5}}$. Yet, with the new slack variables, we have also introduced new semi-infinite constraints $\mbox{C10b}$, $\mbox{C12a}$, and $\mbox{C12b}$. To overcome the related difficulties, in the next subsection, we employ the S-procedure to transform constraints $\mbox{C10b}$, $\mbox{C12a}$, and $\mbox{C12b}$ into equivalent linear matrix inequality constraints, respectively.
\subsection{Handling Semi-Infinite Constraints $\mbox{C10b}$, $\mbox{C12a}$, and $\mbox{C12b}$} 
To tackle the semi-infinite constraints in $\mbox{C10b}$, $\mbox{C12a}$, and $\mbox{C12b}$, we first present the following lemma.
\par
\textit{Lemma~1~(S-Procedure \cite{boyd2004convex}}) Let functions $u_i(\mathbf{p})$, $i\in \left \{ 1,2 \right \}$, $\mathbf{p}\in \mathbb{C}^{N\times 1}$, be defined as
\begin{equation}
u_i(\mathbf{p})= \mathbf{p}^H\mathbf{Q}_i\mathbf{p}+2\Re\left \{\mathbf{q}^H_i\mathbf{p}  \right \}+q_i,\\[-3mm]
\end{equation}
where $\mathbf{Q}_i\in \mathbb{H}^N$, $\mathbf{q}_i\in \mathbb{C}^{N\times 1}$, and $\mathrm{q}_i\in \mathbb{R}$. Then, the implication $u_1(\mathbf{p})\leq0 \Rightarrow u_2(\mathbf{p})\leq0$ holds if and only if a $\delta \geq 0$ exists, such that
\begin{equation}
\delta\hspace{-1mm}\begin{bmatrix}
\mathbf{Q}_1 &  \mathbf{q}_1\\
\mathbf{q}_1^H &  \mathit{q}_1
\end{bmatrix}-\begin{bmatrix}
\mathbf{Q}_2 &  \mathbf{q}_2\\
\mathbf{q}_2^H &  \mathit{q}_2
\end{bmatrix}\succeq \mathbf{0},
\end{equation}
provided that a point $\widehat{\mathbf{p}}$ exists such that $u_i(\widehat{\mathbf{p}})<0$.
\par
To facilitate the application of the S-procedure, we recast constraint $\mbox{C10b}$ as follows
\begin{eqnarray}
&&\hspace*{-10mm}\mbox{C10b:}\hspace*{1mm}\mathbf{\Delta}\mathbf{h}_k^H[m]\Big(\mathbf{W}_k[m]-\lambda_k[m]\big(\underset{\substack{r\in\mathcal{K}\setminus\left\{k\right\} }}{\sum}\mathbf{W}_k[m]+\mathbf{V}[m]\big)\Big)\mathbf{\Delta}\mathbf{h}_k[m]-\lambda_k[m]\sigma_{\mathrm{U}_k}^2\notag\\
&&+2\Re\left\{\overline{\mathbf{h}}_k^H\Big(\mathbf{W}_k[m]-\lambda_k[m]\big(\underset{\substack{r\in\mathcal{K}\setminus\left\{k\right\}}}{\sum}\mathbf{W}_k[m]+\mathbf{V}[m]\big)\Big)\mathbf{\Delta}\mathbf{h}_k[m]\right\}\notag\\
&&+\overline{\mathbf{h}}_k^H\Big(\mathbf{W}_k[m]-\lambda_k[m]\big(\underset{\substack{r\in\mathcal{K}\setminus\left\{k\right\}}}{\sum}\mathbf{W}_k[m]+\mathbf{V}[m]\big)\Big)\overline{\mathbf{h}}_k\geq 0,\hspace*{1mm}\Delta \mathbf{h}_k[m]\in\Omega _k[m].
\end{eqnarray}
\par
Then, by exploiting Lemma 1, constraint $\mbox{C10b}$ can be rewritten as follows
\begin{eqnarray}
\mbox{C10b}
\Leftrightarrow\hspace*{-6mm}&&\tau_k[m]\begin{bmatrix}
 \mathbf{I}_{N_\mathrm{T}}& \mathbf{0}\\
\mathbf{0} & -\mu^2_k[m]\end{bmatrix}-\begin{bmatrix}
 \widehat{\mathbf{W}}_k[m]& \widehat{\mathbf{W}}_k[m]\overline{\mathbf{h}}_k\\
\overline{\mathbf{h}}_k^H\widehat{\mathbf{W}}_k[m] & \overline{\mathbf{h}}_k^H\widehat{\mathbf{W}}_k[m]\overline{\mathbf{h}}_k -\lambda_k[m]\sigma_{\mathrm{U}_k}^2\end{bmatrix}\succeq \mathbf{0}, \notag\\
\Leftrightarrow\hspace*{-6mm}&&\overline{\mbox{C10}\mbox{b}}\mbox{:}\hspace*{1mm}\mathbf{B}_k[m]-\mathbf{C}^H_k\widehat{\mathbf{W}}_k[m]\mathbf{C}_k\succeq\mathbf{0},\forall m,k,
\end{eqnarray}
where $\tau_k[m]\geq 0$ and $\widehat{\mathbf{W}}_k[m]$ is defined as $\widehat{\mathbf{W}}_k[m]\overset{\Delta }{=}\lambda_k[m]\big(\underset{\substack{r\in\mathcal{K}\setminus\left\{k\right\}}}{\sum}\mathbf{W}_r[m]+\mathbf{V}[m]\big)-\mathbf{W}_k[m]$. Besides, matrices $\mathbf{B}_k[m]$ and $\mathbf{C}_k$ are defined as follows
\begin{eqnarray}
\mathbf{B}_k[m]=\begin{bmatrix}
 \tau_k[m]\mathbf{I}_{N_\mathrm{T}}& \mathbf{0}\\
\mathbf{0} & -\tau_k[m]\mu^2_k[m]+\lambda_k[m]\sigma_{\mathrm{U}_k}^2\end{bmatrix},\hspace*{1mm}\forall m,\hspace*{1mm}k,\hspace*{2mm}\mathbf{C}_k=\Big[\mathbf{I}_{N_\mathrm{t}} \hspace*{2mm}\overline{\mathbf{h}}_k\Big],\hspace*{1mm}\forall k.
\end{eqnarray}
\par
Similarly, we resort to Lemma 1 and transform constraints $\mbox{C12a}$ and $\mbox{C12b}$ as follows
\begin{eqnarray}
\hspace*{-6mm}\mbox{C12a}
&\hspace*{-2mm}\Leftrightarrow\hspace*{-2mm}&\iota_{k,j}[m]\begin{bmatrix}
 \mathbf{I}_{N_\mathrm{T}}& \mathbf{0}\\
\mathbf{0} & -\nu_j^2[m]\end{bmatrix}-\begin{bmatrix}
\widetilde{\mathbf{W}}_k[m]& \widetilde{\mathbf{W}}_k[m]\overline{\mathbf{g}}_j\\
\overline{\mathbf{g}}_j^H\widetilde{\mathbf{W}}_k[m] & \overline{\mathbf{g}}_j^H\widetilde{\mathbf{W}}_k[m]\overline{\mathbf{g}}_j +\zeta_{k,j}[m]\end{bmatrix}\succeq \mathbf{0}, \notag\\
&\hspace*{-2mm}\Leftrightarrow\hspace*{-2mm}&\overline{\mbox{C12}\mbox{a}}\mbox{:}\hspace*{1mm}\mathbf{D}_{k,j}[m]-\mathbf{E}^H_j\widetilde{\mathbf{W}}_k[m]\mathbf{E}_j\succeq\mathbf{0},\hspace*{1mm}\forall m,\hspace*{1mm}k,\hspace*{1mm}j,\\
\hspace*{-6mm}\mbox{C12b}
&\hspace*{-2mm}\Leftrightarrow\hspace*{-2mm}&\overline{\mbox{C12}\mbox{b}}\mbox{:}\hspace*{1mm}\varrho _{k,j}[m]\begin{bmatrix}
1& 0\\
0 & -D_j^2[m]\end{bmatrix}-\begin{bmatrix}
\widetilde{\kappa}_{k,j}[m]& \widetilde{\kappa}_{k,j}[m]\overline{d}_j\\
\widetilde{\kappa}_{k,j}[m]\overline{d}_j & \widetilde{\kappa}_{k,j}[m]\overline{d}_j^2- \zeta_{k,j}[m]\end{bmatrix}\succeq \mathbf{0},\hspace*{1mm}\forall m,\hspace*{1mm}k,\hspace*{1mm}j,
\end{eqnarray}
where $\iota_{k,j}[m]$, $\varrho _{k,j}[m]\geq 0$ are slack variables and $\widetilde{\mathbf{W}}_k[m]$ is defined as $\widetilde{\mathbf{W}}_k[m]\overset{\Delta }{=}\mathbf{W}_k[m]-\kappa_k[m]\mathbf{V}[m]$. Also, matrices $\mathbf{D}_{k,j}[m]$ and $\mathbf{E}_j$ are defined as follows
\begin{eqnarray}
\mathbf{D}_{k,j}[m]=\begin{bmatrix}
\iota_{k,j}[m]\mathbf{I}_{N_\mathrm{T}}& \mathbf{0}\\
\mathbf{0} & -\iota_{k,j}[m]\nu_j^2[m]+\zeta_{k,j}[m]\end{bmatrix},\hspace*{1mm}\forall m,k,j,\hspace*{2mm}
\mathbf{E}_j=\Big[\mathbf{I}_{N_\mathrm{T}} \hspace*{4mm}\overline{\mathbf{g}}_j\Big],\hspace*{1mm}\forall j.
\end{eqnarray}
Besides, $\widetilde{\kappa}_{k,j}[m]\in\mathbb{R}$ is defined as $\widetilde{\kappa}_{k,j}[m]=-\kappa_k[m]\sigma_{E_j}^2\alpha^{-1}(1+\rho_j)$ for brevity.
\par
Now, the robust design problem in \eqref{prob3} can be equivalently rewritten as follows
\begin{eqnarray}
\label{prob4}
&&\hspace*{-12mm}\underset{\substack{t[m],\mathbf{W}_{k}[m],\mathbf{V}[m]\in\mathbb{H}^{N_{\mathrm{T}}},\\\xi_k[m],\lambda_k[m],\eta_k[m],\kappa_k[m],\zeta_{k,j}[m],\\ \tau_k[m],\iota_{k,j}[m],\varrho _{k,j}[m]\geq 0}}{\maxo} \,\, \,\, F(t[m],\xi_k[m],\eta_k[m])\overset{\Delta }{=}\frac{1}{T_{\mathrm{tot}}}\underset{m\in\mathcal{M} }{\sum}t[m]\underset{k\in\mathcal{K}}{\sum}\Big(\xi_k[m]-\eta_k[m]\Big)\notag\\
&&\hspace*{2mm}\st\hspace*{17mm}
\mbox{C1-C3},\overline{\mbox{C4}},\overline{\mbox{C5}},\mbox{C6-C9},\mbox{C10a},\overline{\mbox{C10}\mbox{b}},\mbox{C11},\overline{\mbox{C12}\mbox{a}},\overline{\mbox{C12}\mbox{b}}.
\end{eqnarray}
We note that although the S-procedure allows us to sidestep the semi-infinite programming problem, the two resulting constraints $\overline{\mbox{C10}\mbox{b}}$ and $\overline{\mbox{C12}\mbox{a}}$ are non-convex due to the coupling between the optimization variables. Nevertheless, in the following subsection, we will show that by dividing the optimization variables into two blocks, we can efficiently tackle the equivalent optimization problem in \eqref{prob4} and obtain a suboptimal solution with guaranteed convergence.
\subsection{BCD-based Algorithm}
In wireless communications, the coupling of optimization variables often leads to intrinsically challenging non-convex design problems, e.g., for user scheduling \cite{8310563}, subcarrier assignment \cite{8648498}, and intelligent reflecting surface-assisted wireless communication \cite{yu2021smart}. In the literature, BCD-based methods have been shown to be effective for finding high-quality suboptimal solutions for such optimization problems in a computational efficient manner \cite{8930608,9183907}. In particular, by dividing the coupled optimization variables into several blocks, the subproblems associated with the different blocks can be solved in an alternating manner. 
\par
Next, we divide the optimization variables into two blocks, i.e., $\left\{t[m],\mathbf{W}_{k}[m],\mathbf{V}[m],\zeta_{k,j}[m]\right\}$ and $\left\{\lambda_k[m],\kappa_k[m],\xi_k[m],\eta_k[m],\zeta_{k,j}[m] \right\}$, and develop a BCD-based algorithm to tackle optimization problem \eqref{prob4}.\footnote{Note that optimization variable $\zeta_{k,j}[m]$ is not coupled with any other optimization variable. Hence, we optimize $\zeta_{k,j}[m]$ in both blocks to preserve the joint optimality between $\zeta_{k,j}[m]$ and the other optimization variables.}
\par
\textit{Block 1}: We first tackle the optimization of snapshot duration $t[m]$, beamforming matrix $\mathbf{W}_k[m]$, covariance matrix $\mathbf{V}[m]$, and $\zeta_{k,j}[m]$ for given $\left\{\lambda_k[m],\kappa_k[m],\xi_k[m],\eta_k[m],\zeta_{k,j}[m] \right\}$. The corresponding optimization problem for $t[m]$, $\mathbf{W}_k[m]$, and $\mathbf{V}[m]$ is given by
\begin{eqnarray}
\label{prob5}
&&\hspace*{-12mm}\underset{\substack{\mathbf{W}_{k}[m],\mathbf{V}[m]\in\mathbb{H}^{N_{\mathrm{T}}},t[m],\zeta_{k,j}[m]\\ \tau_k[m],\iota_{k,j}[m],\varrho _{k,j}[m]\geq 0}}{\maxo} \,\, \,\, \frac{1}{T_{\mathrm{tot}}}\underset{m\in\mathcal{M} }{\sum}t[m]\underset{k\in\mathcal{K}}{\sum}(\xi_k[m]-\eta_k[m])\notag\\
&&\hspace*{1mm}\st\hspace*{14mm}
\mbox{C1-C3},\overline{\mbox{C4}},\overline{\mbox{C5}},\mbox{C6-C9},\overline{\mbox{C10}\mbox{b}},\overline{\mbox{C12}\mbox{a}},\overline{\mbox{C12}\mbox{b}}.
\end{eqnarray}
We note that in Problem \eqref{prob5}, the objective function and constraints $\overline{\mbox{C4}}$ and $\overline{\mbox{C5}}$ are linear functions and constraints $\overline{\mbox{C10}\mbox{b}}$ and $\overline{\mbox{C12}\mbox{a}}$ are linear matrix inequality (LMI) constraints. The only non-convexity in \eqref{prob5} originates from rank constraint C9. To tackle this problem, we employ SDP relaxation by removing constraint C9 from the problem formulation. The resulting relaxed version of \eqref{prob5} is a convex problem and can be solved efficiently by standard convex optimization solvers such as CVX \cite{grant2008cvx}. Furthermore,
the tightness of the SDP relaxation of optimization problem
\eqref{prob5} is unveiled in the following theorem.
\par
\textit{Theorem 1:}\hspace*{1mm}The optimal beamforming matrix of the relaxed \eqref{prob5} satisfies $\mathrm{Rank}(\mathbf{W}^*_k[m])\leq 1$, $\forall k$, $\forall m$.
\par
\textit{Proof:} The relaxed version of problem \eqref{prob5} has a similar structure as \cite [Problem (43)]{9183907} and Theorem 1 can be proved following the same steps as in \cite [Appendix A]{9183907}. The detailed proof of Theorem 1 is omitted due to page limitation.
\par
Theorem 1 demonstrates that we can always obtain a rank-constrained optimal solution for problem \eqref{prob5}.
Besides, according to Theorem 1, the optimal beamforming
vector $\mathbf{w}^*_k[m]$ can be recovered from $\mathbf{W}^*_k[m]$ by Cholesky decomposition, i.e., $\mathbf{W}^*_k[m]=\mathbf{w}^*_k[m](\mathbf{w}^*_k[m])^H$.
\par
\textit{Block 2}: For $\left\{t[m],\mathbf{W}_k[m],\mathbf{V}[m],\zeta_{k,j}[m]\right\}$, the block $\left\{\lambda_k[m],\kappa_k[m],\xi_k[m],\eta_k[m],\zeta_{k,j}[m] \right\}$ can be obtained by solving the following optimization problem
\begin{eqnarray}
\label{prob6}
&&\hspace*{-12mm}\underset{\substack{\lambda_k[m],\kappa_k[m],\xi_k[m],\eta_k[m],\zeta_{k,j}[m],\\ \tau_k[m],\iota_{k,j}[m],\varrho _{k,j}[m]\geq 0}}{\maxo} \,\, \,\, \underset{m\in\mathcal{M} }{\sum}t[m]\underset{k\in\mathcal{K}}{\sum}(\xi_k[m]-\eta_k[m])\notag\\
&&\hspace*{2mm}\st\hspace*{16mm}
\overline{\mbox{C4}},\overline{\mbox{C5}},\mbox{C10a},\overline{\mbox{C10}\mbox{b}},\mbox{C11},\overline{\mbox{C12}\mbox{a}},\overline{\mbox{C12}\mbox{b}}.
\end{eqnarray}
Here, the objective function and constraints $\overline{\mbox{C4}}$ and $\overline{\mbox{C5}}$ are linear functions and constraint $\mbox{C10a}$ is a convex function. Moreover, given $\mathbf{W}_k[m]$ and $\mathbf{V}[m]$, constraints $\overline{\mbox{C10}\mbox{b}}$ and $\overline{\mbox{C12}\mbox{a}}$ are also LMI constraints. Yet, the optimization problem in \eqref{prob6} is still non-convex due to the logarithmic function in constraint $\mbox{C11}$. Since constraint $\mbox{C11}$ can be rewritten in the form of a difference of convex functions, we propose to employ the IA method \cite{marks1978general} to linearize the non-convex term. To facilitate the application of IA, we construct a global underestimator for the non-convex term based on its first-order Taylor approximation. In particular, we have
\begin{equation}
    \mathrm{log}_2\big(1+\kappa_k[m]\big)\geq \mathrm{log}_2\big(1+\kappa^{(i)}_k[m]\big)+\frac{1}{\mathrm{ln}(2)\big(1+\kappa_k^{(i)}[m]\big)}\big(\kappa_k[m]-\kappa_k^{(i)}[m]\big),\hspace*{1mm}\forall m,k,
\end{equation}
where $\kappa^{(i)}_k[m]$ is the intermediate solution obtained in the $i$-th iteration of the IA and superscript $i$ denotes the iteration index of the optimization variable. As a result, we replace $\mbox{C11}$ by a convex subset which is given by
\begin{eqnarray}
\overline{\mbox{C11}}\mbox{:}\hspace*{1mm}\eta_k[m]-\mathrm{log}_2\big(1+\kappa^{(i)}_k[m]\big)-\frac{1}{\mathrm{ln}(2)\big(1+\kappa_k^{(i)}[m]\big)}\big(\kappa_k[m]-\kappa_k^{(i)}[m]\big)\geq 0,\hspace*{1mm}\forall m,k,
\end{eqnarray}
such that $\overline{\mbox{C11}}\Rightarrow \mbox{C11}$.
\par
\begin{algorithm}[t]
\caption{Inner Approximation-based Algorithm}
\begin{algorithmic}[1]
\small
\STATE Set iteration index $i=1$, error tolerance $0<\epsilon_{\mathrm{IA}}\ll1$, and initial point $\lambda_k^{(i)}[m]$, $\kappa_k^{(i)}[m]$, $\xi_k^{(i)}[m]$, $\eta_k^{(i)}[m]$, $\zeta_{k,j}^{(i)}[m]$, $\forall k\in\mathcal{K}$, $\forall j\in\mathcal{J}$, $\forall m\in\mathcal{M}$
\REPEAT
\STATE For given $\lambda_k^{(i)}[m]$, $\kappa_k^{(i)}[m]$, $\xi_k^{(i)}[m]$, $\eta_k^{(i)}[m]$, $\zeta_{k,j}^{(i)}[m]$, obtain the intermediate solution $\lambda_k^{(i+1)}[m]$, $\kappa_k^{(i+1)}[m]$, $\xi_k^{(i+1)}[m]$, $\eta_k^{(i+1)}[m]$, and $\zeta_{k,j}^{(i+1)}[m]$, $\forall k\in\mathcal{K}$, $\forall j\in\mathcal{J}$, $\forall m\in\mathcal{M}$, by solving optimization problem \eqref{prob7}
\STATE Set $i=i+1$
\UNTIL $\frac{f(\xi_k^{(i)}[m],\eta_k^{(i)}[m])-f(\xi_k^{(i-1)}[m],\eta_k^{(i-1)}[m])}{f(\xi_k^{(i-1)}[m],\eta_k^{(i-1)}[m])}\leq \epsilon_{\mathrm{IA}}$
\end{algorithmic}
\end{algorithm}
\par
Therefore, the optimization problem to be solved in the $(i+1)$-th iteration of the IA-based algorithm is given as follows
\begin{eqnarray}
\label{prob7}
&&\hspace*{-12mm}\underset{\substack{\lambda_k[m],\kappa_k[m],\xi_k[m],\eta_k[m],\zeta_{k,j}[m],\\ \tau_k[m],\iota_{k,j}[m],\varrho _{k,j}[m]\geq 0}}{\maxo} \,\, \,\, f(\xi_k[m],\eta_k[m])\overset{\Delta }{= }\frac{1}{T_{\mathrm{tot}}}\underset{m\in\mathcal{M} }{\sum}t[m]\underset{k\in\mathcal{K}}{\sum}(\xi_k[m]-\eta_k[m])\notag\\
&&\hspace*{2mm}\st\hspace*{16mm}
\overline{\mbox{C4}},\overline{\mbox{C5}},\mbox{C10a},\overline{\mbox{C10}\mbox{b}},\overline{\mbox{C11}},\overline{\mbox{C12}\mbox{a}},\overline{\mbox{C12}\mbox{b}}.
\end{eqnarray}
The proposed IA-based algorithm is summarized in \textbf{Algorithm 1}. Note that in each iteration of \textbf{Algorithm 1}, we solve a convex optimization problem and corresponding the objective function $f(\xi_k[m],\eta_k[m])$ is monotonically non-decreasing. Moreover, according to \cite[Theorem 1]{marks1978general}, the proposed IA-based algorithm is guaranteed to converge to a locally optimal solution of \eqref{prob6} in polynomial time.
\par
The overall BCD-based algorithm is summarized in \textbf{Algorithm 2}. Recall that the objective function in \eqref{prob7} is monotonically non-decreasing in each iteration of \textbf{Algorithm 1} and \eqref{prob5} is a standard semidefinite programming problem. As a result, in each iteration, the objective function value of the optimization problem in \eqref{prob4} does not decrease. Thus, according to \cite{razaviyayn2013unified}, the proposed BCD-based algorithm is guaranteed to converge to a stationary point of optimization problem \eqref{prob2}. Due to the application of the bound on the uncertainty region derived in Section \ref{ISAC_safe_approximation}, the obtained solution is a feasible suboptimal solution of the original problem in \eqref{prob1}. Furthermore, in the IA-based algorithm, we solve for a set of scalar optimization variables, i.e., $\left\{\lambda_k[m],\kappa_k[m],\xi_k[m],\eta_k[m],\zeta_{k,j}[m] \right\}$, while the optimization problem in \eqref{prob5} involves positive semidefinite matrices $\mathbf{W}_k[m]$ and $\mathbf{V}[m]$. According to \cite{arora2009computational}, the computational complexity of the overall algorithm is dominated by step 3 of \textbf{Algorithm 2}. In particular, the computational complexity needed to solve an SDP problem with a set of $a$ SDP constraints, where each constraint contains a $b\times b$ positive semidefinite matrix, is given by $\mathcal{O}\big(ab^3+a^2b^2+a^3\big)$ \cite[Theorem 3.12]{polik2010interior}. For the rank-relaxed version of problem \eqref{prob5}, we have $a=MK(J+1)$ and $b=N_{\mathrm{T}}$. Therefore, the overall computational complexity of the proposed BCD algorithm is given by $\mathcal{O}\Big(\mathrm{log}(\frac{1}{\epsilon_{\mathrm{BCD}}})\big[MK(J+1)N_{\mathrm{T}}^3+\big(MK(J+1)\big)^2N_{\mathrm{T}}^2+\big(MK(J+1)\big)^3\big]\Big)$, where $\epsilon_{\mathrm{BCD}}$ is the pre-defined accuracy factor in \textbf{Algorithm 2}.\footnote{Due to the random initialization and channel realizations as well as the unknown structure of the feasible region, it is very challenging to analyze the speed of convergence of the proposed algorithm as a function of $M$ and $N_{\mathrm{T}}$. Therefore, in Section \ref{ISAC_simulation}, we investigate the speed of convergence of the proposed BCD-based algorithm via computer simulations.}
\begin{algorithm}[t]
\caption{Block Coordinate Descent-based Algorithm}
\begin{algorithmic}[1]
\small
\STATE Set iteration index $l=1$, convergence tolerance $0\leq\epsilon_{\mathrm{BCD}}\ll1$, randomly generate $(\mathbf{W}_k[m])^1$ and $(\mathbf{V}[m])^1$ by drawing their elements uniformly from the interval $(0, \frac{P^{\mathrm{max}}}{(K+1)N_{\mathrm{T}}})$, and set $(t_k[m])^1=\frac{T_{\mathrm{tot}}}{M}$, $(\lambda_k[m])^1=2^{\overline{R}_{\mathrm{req}_k}}-1$, $(\kappa_k[m])^1=2^{\overline{R}_{\mathrm{tol}_k}}-1$, $(\xi_k[m])^1=\overline{R}_{\mathrm{req}_k}$, $(\eta_k[m])^1=\overline{R}_{\mathrm{tol}_k}$, $(\zeta_{k,j}[m])^1=0$, $\forall k\in\mathcal{K}$, $ \forall j\in\mathcal{J}$, $\forall m\in\mathcal{M}$
\REPEAT
\STATE Solve \eqref{prob5} for given $(\lambda_k[m])^l$, $(\kappa_k[m])^l$, $(\xi_k[m])^l$, $(\eta_k[m])^l$, $\zeta_{k,j}[m]^l$ and obtain $(t[m])^{l+1}$, $(\mathbf{W}_k[m])^{l+1}$, $(\mathbf{V}[m])^{l+1}$, $(\zeta_{k,j}[m])^{l+1}$
\STATE Solve \eqref{prob7} for $t[m]=(t[m])^{l+1}$, $\mathbf{W}_k[m]=(\mathbf{W}_k[m])^{l+1}$, $\mathbf{V}[m]=(\mathbf{V}[m])^{l+1}$, and $\zeta_{k,j}[m]=(\zeta_{k,j}[m])^{l+1}$ by applying \textbf{Algorithm 1} and obtain $(\lambda_k[m])^{l+1}$, $(\kappa_k[m])^{l+1}$, $(\xi_k[m])^{l+1}$, $(\eta_k[m])^{l+1}$, $\zeta_{k,j}[m]^{l+1}$
\STATE Set $l=l+1$
\UNTIL $\frac{ F\Big((t[m])^l,(\xi_k[m])^l,(\eta_k[m])^l)\Big)}{F\Big((t[m])^{l-1},(\xi_k[m])^{l-1},(\eta_k[m])^{l-1})\Big)}-1\leq \epsilon_{\mathrm{BCD}}$, where $F(\cdot,\cdot,\cdot)$ is defined in \eqref{prob4}
\STATE $t^*[m]=(t[m])^l$, $\mathbf{W}_k^*[m]=(\mathbf{W}_k^*[m])^l$, $\mathbf{V}^*[m]=(\mathbf{V}^*[m])^l$
\end{algorithmic}
\end{algorithm}

\section{Simulation Results}
\label{ISAC_simulation}
In this section, the system performance of the
proposed resource allocation scheme is evaluated via simulations. 
\subsection{Simulation Setup}
We focus on the resource allocation in an ISAC system covering a $120$-degree sector of a cell with a radius of $200$ m. In particular, the DFRC BS is equipped with $N_{\mathrm{T}}=12$ antennas with half-wavelength spacing of neighboring antenna elements. There are $K$ legitimate users and $J$ previously detected targets (potential eavesdroppers) randomly and uniformly distributed in the sector. Given the scanning period $T_{\mathrm{tot}}$, the DFRC BS sequentially scans the sector using $M$ consecutive snapshots for new and already known targets where each snapshot is associated with the transmission of a dedicated beam pattern pointing in one specific direction of the sector. The $M$ highly-directional beam patterns, $\mathbf{R}_d[m]$, are generated in a similar manner as in \cite[Section IV-B]{4516997}, where the main lobe of each beam pattern has a beamwidth of $\frac{120}{M}$ degrees. Hence, for a given $P^{\mathrm{max}}$, a set of $\mathbf{R}_d[m]$, $\forall m\in\mathcal{M}$, is pre-designed for the considered resource allocation optimization problem in \eqref{prob2}. Moreover, we model the small-scale fading coefficients of the channels of the legitimate users as independent and identically distributed Rayleigh random variables. The estimated LoS component of the channels of the potential eavesdroppers, i.e., $\overline{\mathbf{g}}_j$, is modeled as the product of the array response vectors of the transceivers with Ricean factor $\rho_j=5$. The path loss at a reference distance of $1$ m is set to $\alpha=46$ dB. To determine the upper bound for the uncertainty regarding the eavesdropper channels $\mathbf{\Delta g}_j[m]$, i.e., $\nu_j[m]$, we set the maximum angle uncertainty to $\phi_j[m]=5$ degrees. Moreover, for a given Ricean factor $\rho_j$, the upper bound for $\widetilde{g}_{j,n}[m]$, i.e., $\beta_{j,n}[m]$, is set to $\beta_{j,n}[m]=0.1\sqrt{\rho_j}$. Based on this, we can obtain the upper bound on $\mathbf{\Delta g}_j[m]$ by applying \eqref{upperbound-eaves_channel1}. The bound of the distance uncertainty of known target $j$ in snapshot $m$ is assumed to be $D_j[m]=5$ m, $\forall j$, $m$. To facilitate the presentation, in the sequel, we define the normalized beam pattern difference tolerance factor and the maximum normalized estimation error of the legitimate user channels as $\varsigma_d[m]=\frac{\delta_d[m]}{\left\|\mathbf{R}_d[m] \right\|^2_F}$ and $\chi_k^2[m]=\frac{\mu_k^2[m]}{\left\|\overline{\mathbf{h}}_k \right\|^2_2}$, respectively. Also, to facilitate the investigation of the average system performance, the aforementioned parameters are assumed to be identical for all snapshots, i.e., $\phi_j[m]=\phi_j$, $\beta_{j,n}[m]=\beta_{j,n}$, $D_j[m]=D_j$, $\varsigma_d[m]=\varsigma_d$, and $\chi_k^2[m]=\chi_k^2$, $\forall m\in\mathcal{M}$. Unless otherwise specified, we adopted the default parameter values introduced above or listed in Table \ref{tab:ISAC_parameters} for our simulations.
\begin{table}[t]\vspace*{0mm}\caption{System simulation parameters.}
\label{tab:ISAC_parameters}\footnotesize
\newcommand{\tabincell}[2]{\begin{tabular}{@{}#1@{}}#2\end{tabular}}
\centering
\begin{tabular}{|l|l|l|}\hline
    \hspace*{-1mm}$f_c$ & Carrier frequency & $5$ GHz\\
\hline
    \hspace*{-1mm}$T_{\mathrm{tot}}$ & Duration of scanning period & $5$ ms \\
\hline
    \hspace*{-1mm}$\sigma_{\mathrm{U}_k}^2$ & Noise power at legitimate user $k$ & $-100$ dBm\\
\hline
    \hspace*{-1mm}$\sigma_{\mathrm{E}_j}^2$ & Noise power at potential eavesdropper $j$ & $-100$ dBm \\
\hline
    \hspace*{-1mm}$t_{\mathrm{min}}$ & Minimum duration of snapshot & $0.1$ ms\\
\hline
    \hspace*{-1mm}$t_{\mathrm{max}}$ & Maximum duration of snapshot & $4$ ms \\
\hline
    \hspace*{-1mm}$\overline{R}_{\mathrm{req}_k}$ & Average achievable rate of users & $0.5$ bits/s/Hz\\
\hline
    \hspace*{-1mm}$\overline{R}_{\mathrm{tol}_k}$ & Maximum tolerable information leakage rate & $0.2$ bits/s/Hz \\
\hline
    \hspace*{-1mm}$\epsilon_{\mathrm{IA}}$ & Convergence tolerance of IA algorithm & $10^{-2}$\\
\hline
    \hspace*{-1mm}$\epsilon_{\mathrm{BCD}}$ & Convergence tolerance of BCD algorithm & $10^{-3}$ \\
\hline
\end{tabular}
\end{table}
\subsection{Baseline Schemes}
For comparison, we consider three baseline schemes. For baseline scheme 1, the scanning period is divided into $M$ equal-length snapshots. Then, we jointly optimize the beamforming vectors and the covariance matrix of the AN for maximization of the system sum secrecy rate in the given scanning period. For baseline scheme 2, we consider a non-robust scheme that treats the estimated CSI as perfect and adopt zero-forcing (ZF) beamforming at the DFRC BS. In particular, the direction of the beamformer $\mathbf{w}_k[m]$ for legitimate user $k$ is fixed and lies in the null space of all other legitimate user channels. Then, we tackle \eqref{prob2} by jointly optimizing the design of the duration of each snapshot, the covariance matrix of the AN, and the transmit power for beamformer $\mathbf{w}_k[m]$. Moreover, to investigate the actual performance of practical ISAC systems, for baseline scheme~2, we substitute the solutions obtained by solving \eqref{prob2} back to the problem in \eqref{prob1} and compute the actual system sum secrecy rate using the objective function of \eqref{prob1}. For baseline scheme 3, we employ the single snapshot multi-stage design approach. In particular, assuming the estimates of the angles of all potential eavesdroppers are known at the DFRC BS, we first design a multi-beam beam pattern $\mathbf{R}_d$ for given $P^{\mathrm{max}}$ by employing the method proposed in \cite[Section IV-B]{4516997}, where the different beams of the multi-beam beam pattern point to the potential eavesdroppers to facilitate high-quality sensing. Then, after setting $M=1$ and $t[1]=T_{\mathrm{tot}}$, we tackle the optimization problem in \eqref{prob2} subject to constraints C1, C2, C4, C5, C7, C8, and C9 by applying \textbf{Algorithm 2} and obtain the desired beamforming vectors and covariance matrix of the AN.\footnote{Note that for baseline scheme 3, we first have to design a beam pattern $\mathbf{R}_d$ for each channel realization, and then, solve a problem similar to \eqref{prob2} by applying \textbf{Algorithm 2}. In fact, this baseline scheme is considerably more time-consuming to simulate due to its high complexity than our proposed scheme. As a result, we only show the performance of baseline scheme 3 in Figure \ref{fig:ISAC_uncertainty}.} Besides, the results in Figures \ref{fig:ISAC_convergence}-\ref{fig:ISAC_uncertainty} are averaged over feasible solutions only. The feasibility of the considered scheme is studied in Figure \ref{fig:ISAC_outageprob}.
\subsection{Convergence of the Proposed BCD Algorithm}
\begin{figure}[t]
\begin{minipage}[b]{0.47\linewidth}
\hspace*{-5mm}\includegraphics[width=3.2in]{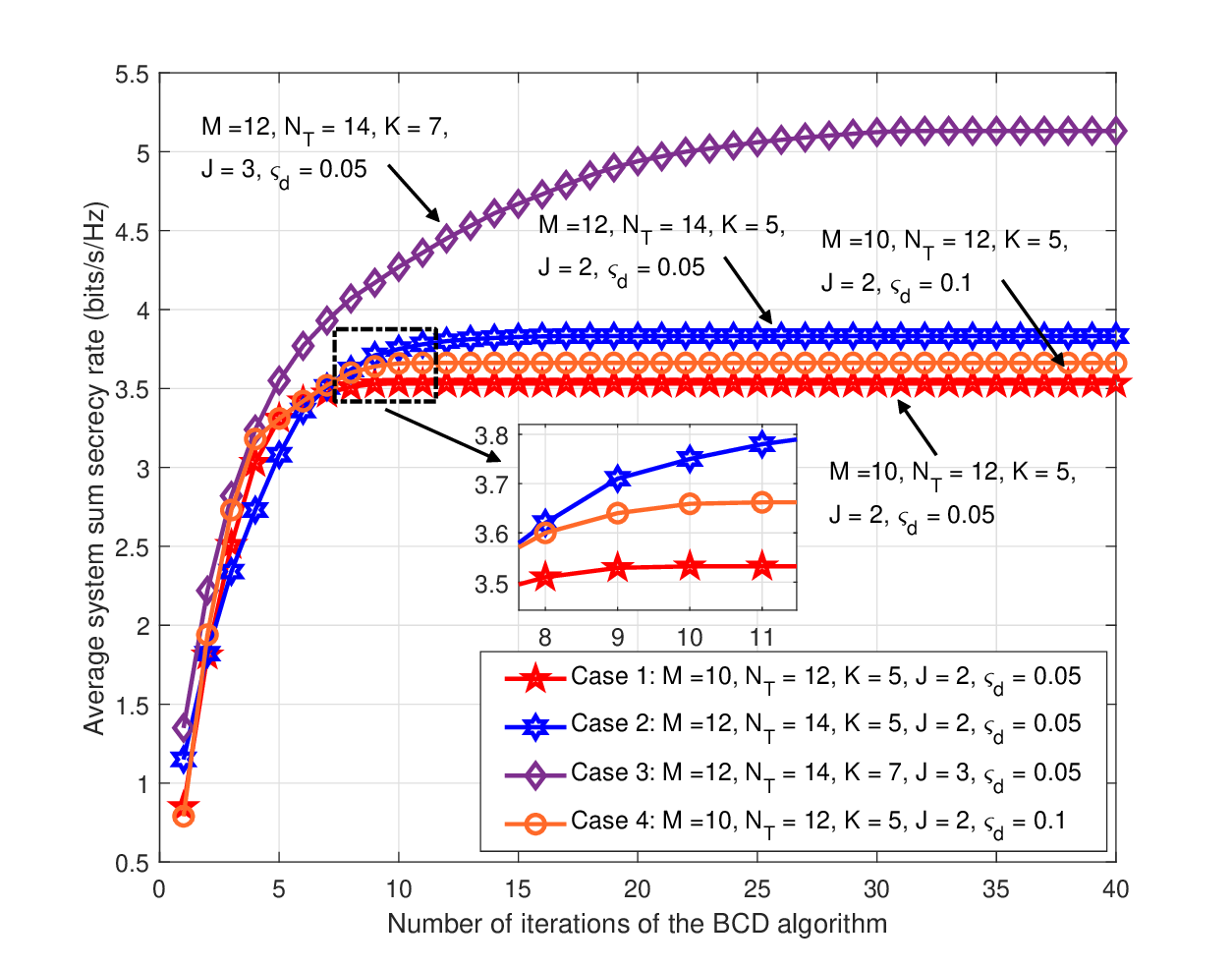}
\caption{Convergence of the proposed BCD-based algorithm for different values of $M$, $N_{\mathrm{T}}$, $K$, and $J$. The system parameters are set as $P^{\mathrm{max}}=30$ dBm, $\varsigma_d=0.05$, $\chi_k^2=0.1$, and $\overline{R}_{\mathrm{req}_k}=0.5$ bits/s/Hz.}
\label{fig:ISAC_convergence}
\end{minipage}\hspace*{8mm}
\begin{minipage}[b]{0.47\linewidth}
\hspace*{-5mm}\includegraphics[width=3.2in]{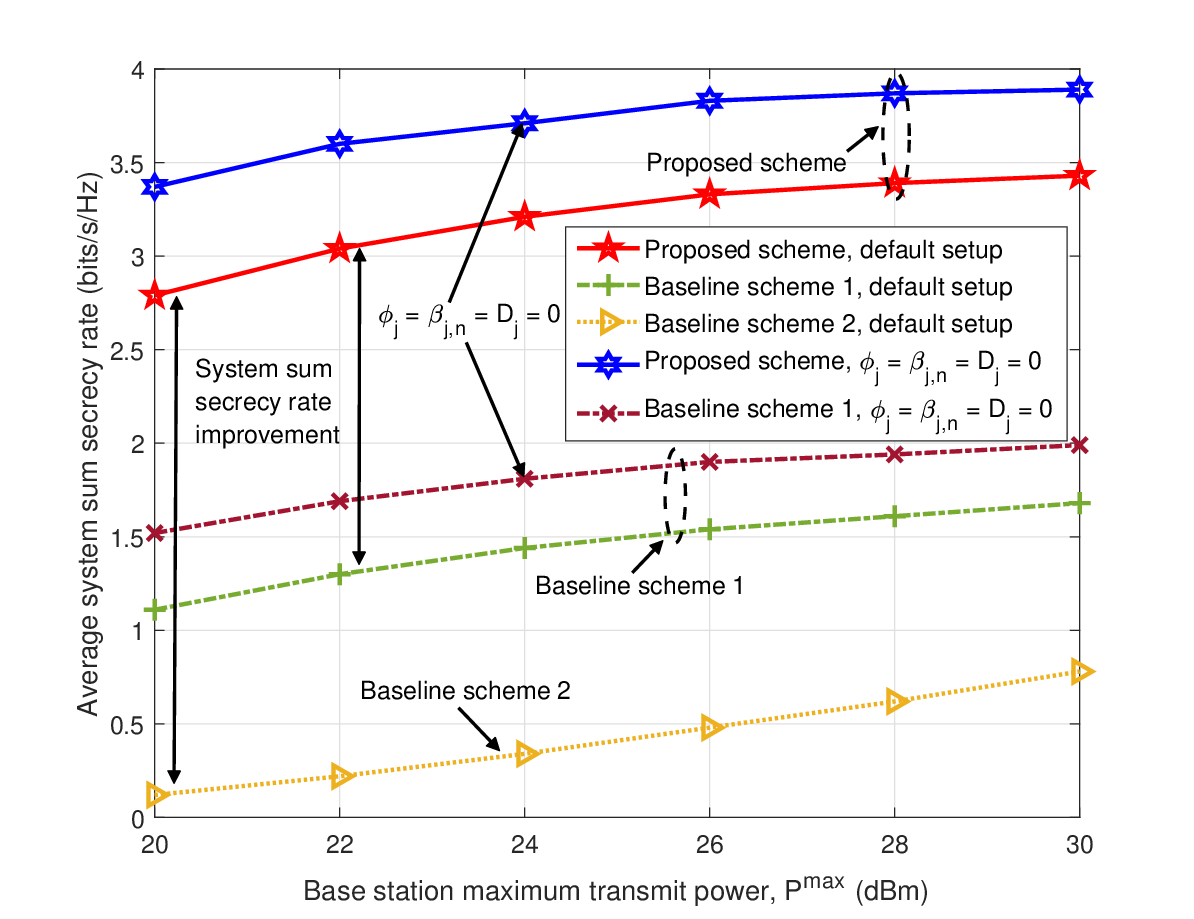}
\caption{Average system sum secrecy rate (bits/s/Hz) versus the maximum transmit power of the DFRC BS, $P^{\mathrm{max}}$, for different schemes with $M=10$, $K=5$, $J=2$, $\varsigma_d=0.05$, $\chi_k^2=0.1$, and $\overline{R}_{\mathrm{req}_k}=0.5$ bits/s/Hz.}
\label{fig:ISAC_power}
\end{minipage}
\end{figure}
In Figure \ref{fig:ISAC_convergence}, we investigate the convergence of the proposed \textbf{Algorithm 2} for different numbers of antenna elements $N_{\mathrm{T}}$, snapshots $M$, legitimate users $K$, and potential eavesdroppers $J$. In particular, we consider four cases: Case 1 with $M=10$, $N_{\mathrm{T}}=12$, $K=5$, $J=2$, and $\varsigma_d[m]=0.05$; Case 2 with $M=12$, $N_{\mathrm{T}}=14$, $K=5$, $J=2$, and $\varsigma_d[m]=0.05$; Case 3 with $M=12$, $N_{\mathrm{T}}=14$, $K=7$, $J=3$, and $\varsigma_d[m]=0.05$; and Case 4 with $M=10$, $N_{\mathrm{T}}=12$, $K=5$, $J=2$, and $\varsigma_d[m]=0.1$. As can be observed, for all four cases, the proposed BCD algorithm monotonically converges quickly to a stationary point. Specifically, for Case 1, the proposed algorithm converges within $10$ iterations of \textbf{Algorithm 2} on average. For Case 2, the proposed algorithm needs roughly 5 additional iterations to converge to a slightly larger objective value compared to Case 1. This is because the larger values for $M$ and $N_{\mathrm{T}}$ facilitate a more efficient beamforming policy and the corresponding feasible region of problem \eqref{prob2} expands with $M$ and $N_{\mathrm{T}}$. Moreover, compared to Case 1 and Case 2, for Case 3, the proposed algorithm needs around $25$ additional iterations for convergence since the larger values of $K$ and $J$ lead to a larger number of optimization variables and constraints. Besides, compared to Case 1, for Case 4, the proposed scheme converges to a slightly larger objective function value in roughly the same number of iterations. This is because, by relaxing the beam pattern mismatch requirement, the feasible set of the considered optimization problem is enlarged. As a result, the proposed algorithm may converge to a more efficient solution, leading to a higher system sum secrecy rate. Overall, we conclude that the proposed algorithm enjoys fast convergence.
\subsection{Average System Sum Secrecy Rate versus Maximum BS Transmit Power}
\begin{figure}[t]
\begin{minipage}[b]{0.47\linewidth}
\hspace*{-5mm}\includegraphics[width=3.2in]{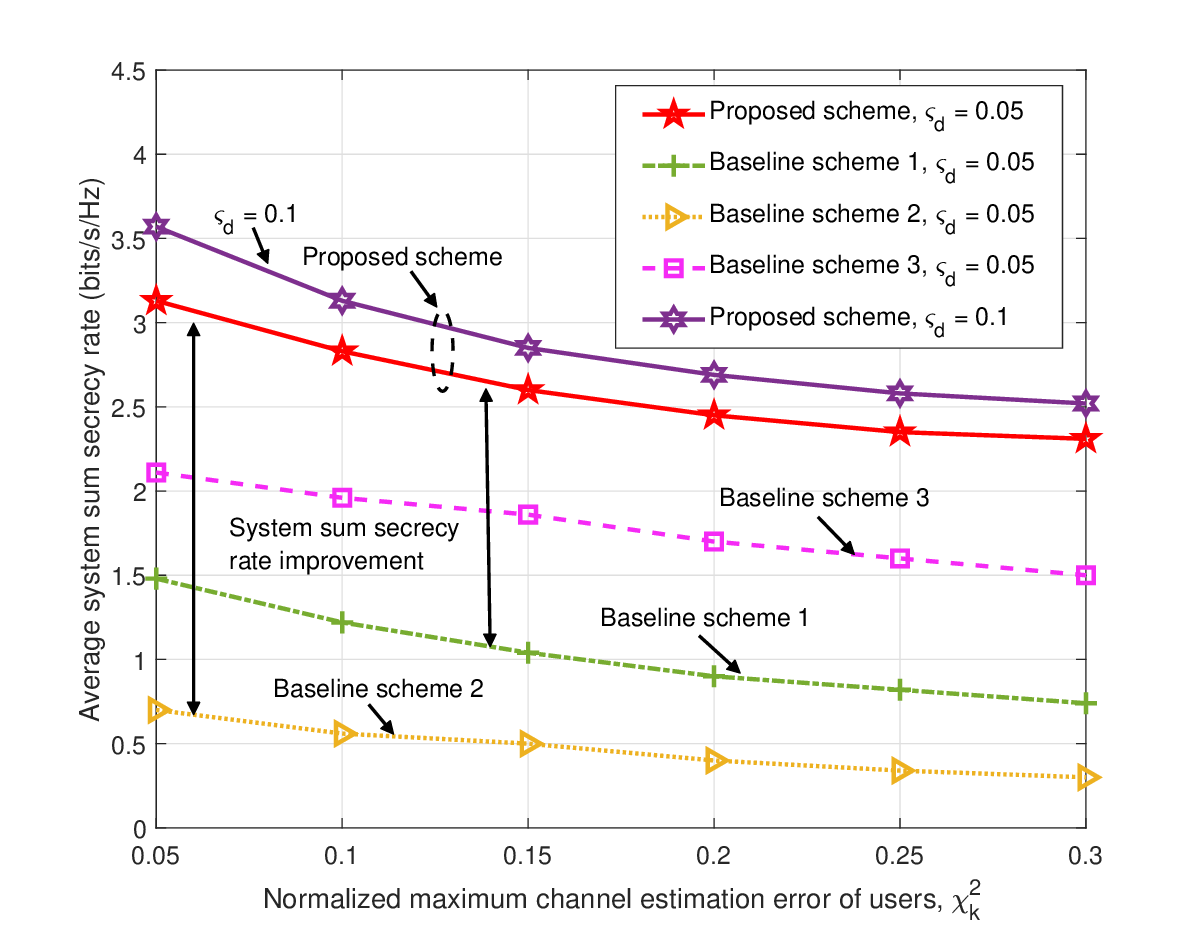}
\caption{Average system sum secrecy rate (bits/s/Hz) versus normalized maximum user CSI uncertainty with $M=10$, $P^{\mathrm{max}}=35$ dBm, $K=5$, $J=4$, and $\overline{R}_{\mathrm{req}_k}=0.4$ bits/s/Hz.}
\label{fig:ISAC_uncertainty}
\end{minipage}\hspace*{8mm}
\begin{minipage}[b]{0.47\linewidth}
\hspace*{-5mm}\includegraphics[width=3.2in]{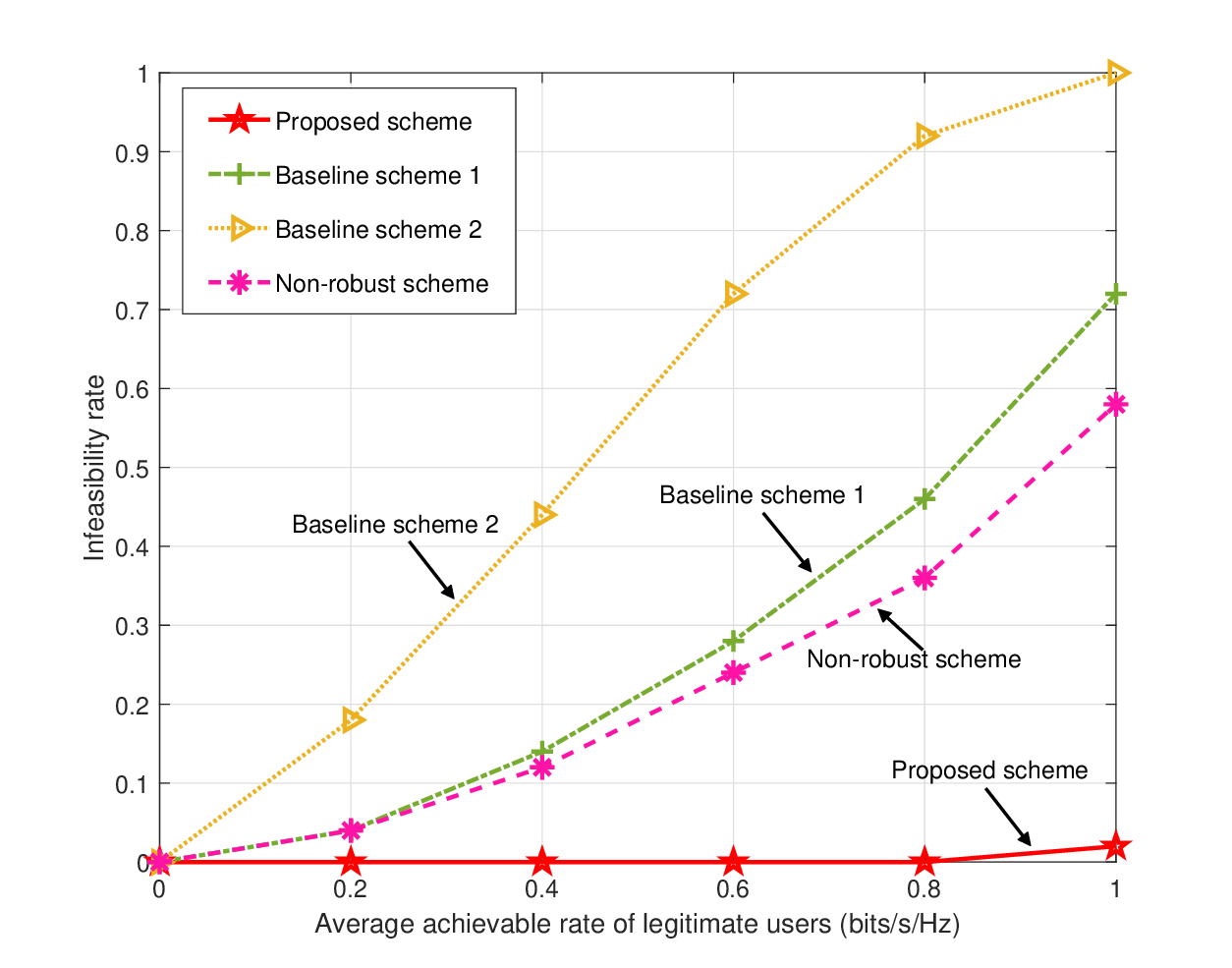}
\caption{Average infeasibility rate versus average achievable rate of the legitimate users, $\overline{R}_{\mathrm{req}_k}$, for different schemes with $M=10$, $P^{\mathrm{max}}=35$ dBm, $K=3$, $J=2$, $\varsigma_d=0.1$, and $\chi_k^2=0.1$.}
\label{fig:ISAC_outageprob}
\end{minipage}
\end{figure}

In Figure \ref{fig:ISAC_power}, we study the average system sum secrecy rate versus the maximum transmit power of the DFRC BS, $P^{\mathrm{max}}$, for different resource allocation schemes. As expected, the system sum secrecy rate increases monotonically with $P^{\mathrm{max}}$. Moreover, we observe that the proposed scheme leads to a substantially higher system sum secrecy rate compared to baseline schemes 1 and 2. This significant performance improvement is due to the joint optimization of the beamforming vectors, the covariance matrix of the AN, and the duration of the snapshots. On the one hand, in each snapshot, the proposed scheme allows the DFRC BS to focus its beam on the legitimate users and impair the eavesdroppers' channels by exploiting AN which facilitates high data-rate secure communications. On the other hand, benefiting from the optimized variable snapshot lengths, the proposed scheme can exploit the extra DoFs introduced by the variable snapshot durations to strike a balance between sensing and communication performance. By contrast, the two baseline schemes achieve a considerably lower system sum secrecy rate. In particular, for baseline scheme 1, although the DFRC BS can still exploit the DoFs available for resource allocation in each snapshot, it is limited by the fixed snapshot durations. As for baseline scheme 2, although it is able to mitigate the MUI via ZF beamforming, it leads to a considerably lower system sum secrecy rate.\footnote{We note that baseline scheme 2 is prone to infeasibility due to constraint C4 in optimization problem \eqref{prob2}, cf. Figure \ref{fig:ISAC_outageprob}.} This can be explained as follows. First, due to the partially fixed beamforming policy and the emission of AN, baseline scheme 2 may not be able to provide high data-rate communication services to all legitimate users. Second, since ZF beamforming does not exploit the CSI of the potential eavesdroppers, it may also lead to high channel capacities between the DFRC BS and the potential eavesdroppers for wiretapping the legitimate users. Besides, compared to the case of perfect CSI of the potential eavesdroppers ($\phi_j=0$, $\beta_{j,n}=0$, and $d_j=0$), thanks to the proposed robust optimization framework, for the proposed scheme, small distance, angle, and multi-path fading uncertainties cause only a small degradation of the system performance. Specifically, for the proposed scheme and baseline scheme 1, in the presence of eavesdropper CSI uncertainty, the DFRC BS allocates more power to the AN to effectively impair the channels of the potential eavesdroppers.
\subsection{Average System Sum Secrecy Rate versus Normalized User CSI Uncertainty}
In Figure \ref{fig:ISAC_uncertainty}, we illustrate the average system sum secrecy rate versus the maximum normalized channel estimation error of the legitimate users. As expected, the average system sum secrecy rate for the proposed scheme and the three baseline schemes decreases for increasing $\chi^2_k$. This is due to the fact that as $\chi^2_k$ increases, the DFRC BS becomes less flexible and more conservative in the joint design of the beamforming vectors, AN, and snapshot duration in order to be able to satisfy the QoS constraints. Specifically, in the presence of CSI uncertainty for the legitimate users, the beamforming for the data transmission has to become less directional, which potentially decreases the SINR of the legitimate users. Moreover, over the entire considered range of $\chi^2_k$, the proposed scheme significantly outperforms the three baseline schemes. This unveils that by jointly exploiting all available DoFs of the ISAC system, the proposed scheme can guarantee physical layer security more effectively than the three baseline schemes, even if the CSI of the legitimate users is not perfectly known. Furthermore, although the conventional multi-stage optimization framework in baseline scheme 3 allows the DFRC BS to potentially serve all the legitimate users in one snapshot, its system sum secrecy rate is compromised by two factors. First, for multiple sensing targets, the desired beam pattern $\mathbf{R}_d$ has to generate multiple beams where each beam aligns with one of the sensing targets. Yet, the resulting multi-beam beam pattern can potentially cause significant information leakage if the AN cannot simultaneously impair the channels of all potential eavesdroppers. This inevitably results in a high channel capacity between the BS and the potential eavesdroppers, which reduces the system sum secrecy rate. Second, for stringent beam pattern mismatch requirements, which are favorable for target sensing, e.g., $\varsigma_d=0.05$, the DFRC BS has to exploit most of the available DoFs to synthesize the desired multi-beam beam pattern. As a result, the flexibility in information beamforming design is small, which results in low SINRs for the legitimate users. Besides, comparing the performance of the proposed scheme for $\varsigma_d=0.05$ and $\varsigma_d=0.1$, the former yields a lower system sum secrecy rate. Recall that the normalized beam pattern difference tolerance factor $\varsigma_d$ limits the deviation of the actual beam pattern from the desired one. As a result, for larger values of $\varsigma_d$, the DFRC BS has more DoFs for beamforming vector optimization, which facilitates a more efficient communication-oriented beam pattern design. The above observations indicate that the system designer has to carefully choose the value of $\varsigma_d$ for ISAC system design.
\subsection{Infeasibility rate versus Average User Achievable Rate}
Figure \ref{fig:ISAC_outageprob} shows the infeasibility rate of optimization problem \eqref{prob1} versus the average achievable rate of the legitimate users for different resource allocation schemes. Infeasibility occurs if a solution that simultaneously satisfies the desired achievable rate $\overline{R}_{\mathrm{req}_k}$ requirement and all the other constraints in optimization problem \eqref{prob1} cannot be found. To this end, we first generate $100$ random channel realizations for the considered ISAC system and solve the approximated problem in \eqref{prob2} for a given $\overline{R}_{\mathrm{req}_k}$. Then, we count the number of feasible solutions of \eqref{prob1}, denoted by $I_{\mathrm{fea}}$, and compute the infeasibility rate as $1-\frac{I_{\mathrm{fea}}}{100}$. As expected, the infeasibility rates of the proposed scheme and baseline schemes 1 and 2 are monotonically non-decreasing as $\overline{R}_{\mathrm{req}_k}$ increases. Yet, compared to the two baseline schemes, the proposed scheme can significantly reduce the infeasibility rate thanks to the proposed optimization framework. In particular, for larger values of $\overline{R}_{\mathrm{req}_k}$, the DFRC BS employing the proposed scheme is able to flexibly prioritize the legitimate users with unfavorable channel conditions by extending the duration of the snapshots of the synthesized beam pattern. For comparison, in addition to the non-robust baseline scheme 2, we also study the infeasibility rate of another non-robust scheme, i.e., the proposed scheme taking CSI uncertainty not into account. Specifically, for this non-robust scheme, we solve an optimization problem similar to \eqref{prob1} but treat the estimated CSI of the legitimate users and angle of the potential eavesdroppers as perfect. Also, the multi-path fading component is assumed to be absent in the potential eavesdroppers channels. The infeasibility rate of this non-robust scheme is computed in a similar way as that of baseline scheme 2. As can be observed from Figure \ref{fig:ISAC_outageprob}, because of the CSI uncertainty in the ISAC system, the non-robust scheme leads to roughly $10\%$ infeasibility rate even for a low average achievable rate constraint of the legitimate users. As $\overline{R}_{\mathrm{req}_k}$ increases, the non-robust scheme yields a very high infeasibility rate compared to the proposed scheme. These results not only underscore the effectiveness of the proposed scheme in facilitating high data-rate communications, but also underline the importance of taking into account the CSI uncertainty of both the legitimate users and the potential eavesdroppers for robust resource allocation algorithm design for ISAC systems.
\vspace*{-2mm}
\subsection{Beam Patterns During Scanning Period}
\begin{figure}[t]
\begin{minipage}[b]{0.47\linewidth}
\hspace*{-5mm}\includegraphics[width=3.2in]{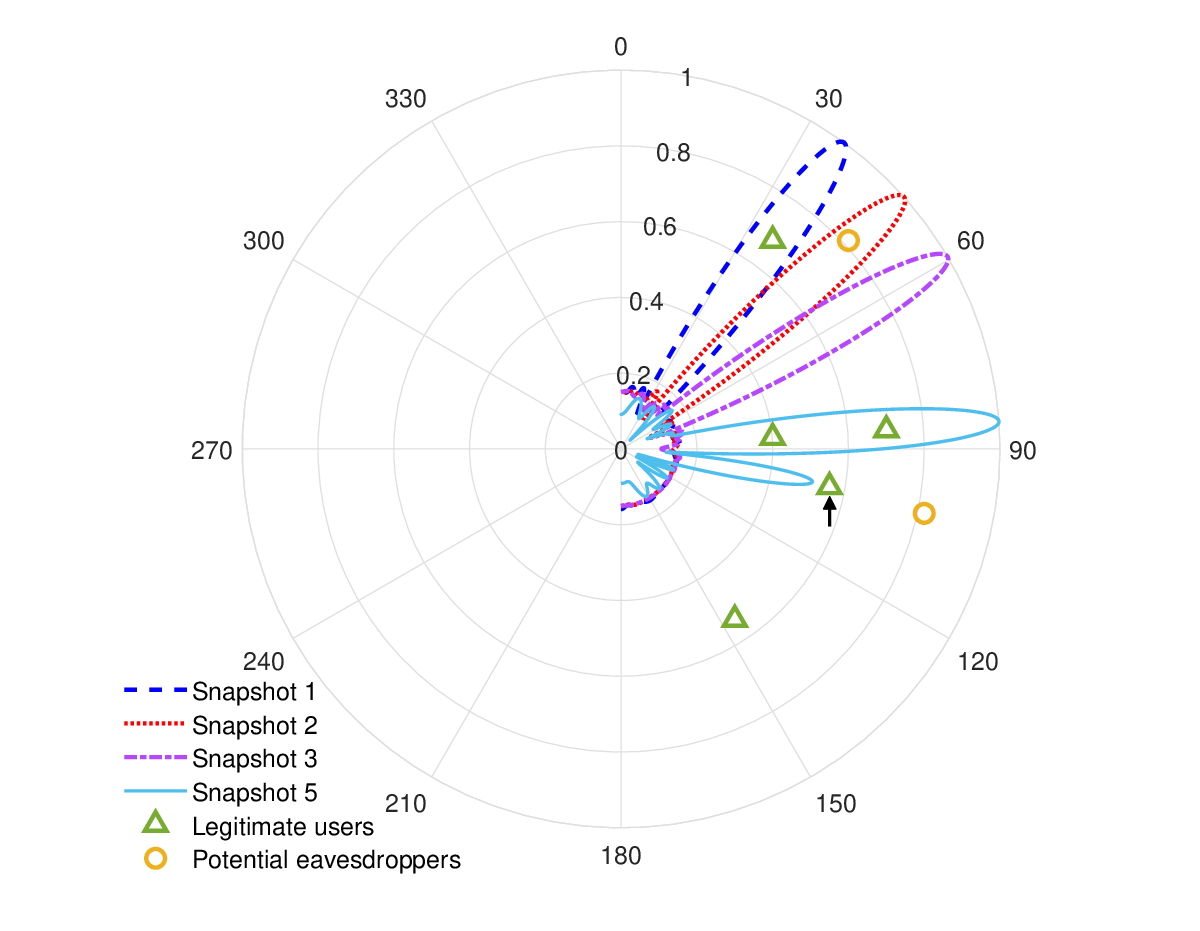}
\caption{Illustration of the considered ISAC system with $M=10$, $K=5$, $J=2$, and the radius of the sector is $200$ m. The polar plot shows the beam patterns for snapshots $1$, $2$, $3$, and $5$ as well as the locations of the legitimate users and potential eavesdroppers.}
\label{fig:ISAC_polar}
\end{minipage}\hspace*{8mm}
\begin{minipage}[b]{0.47\linewidth}
\hspace*{-5mm}\includegraphics[width=3.2in]{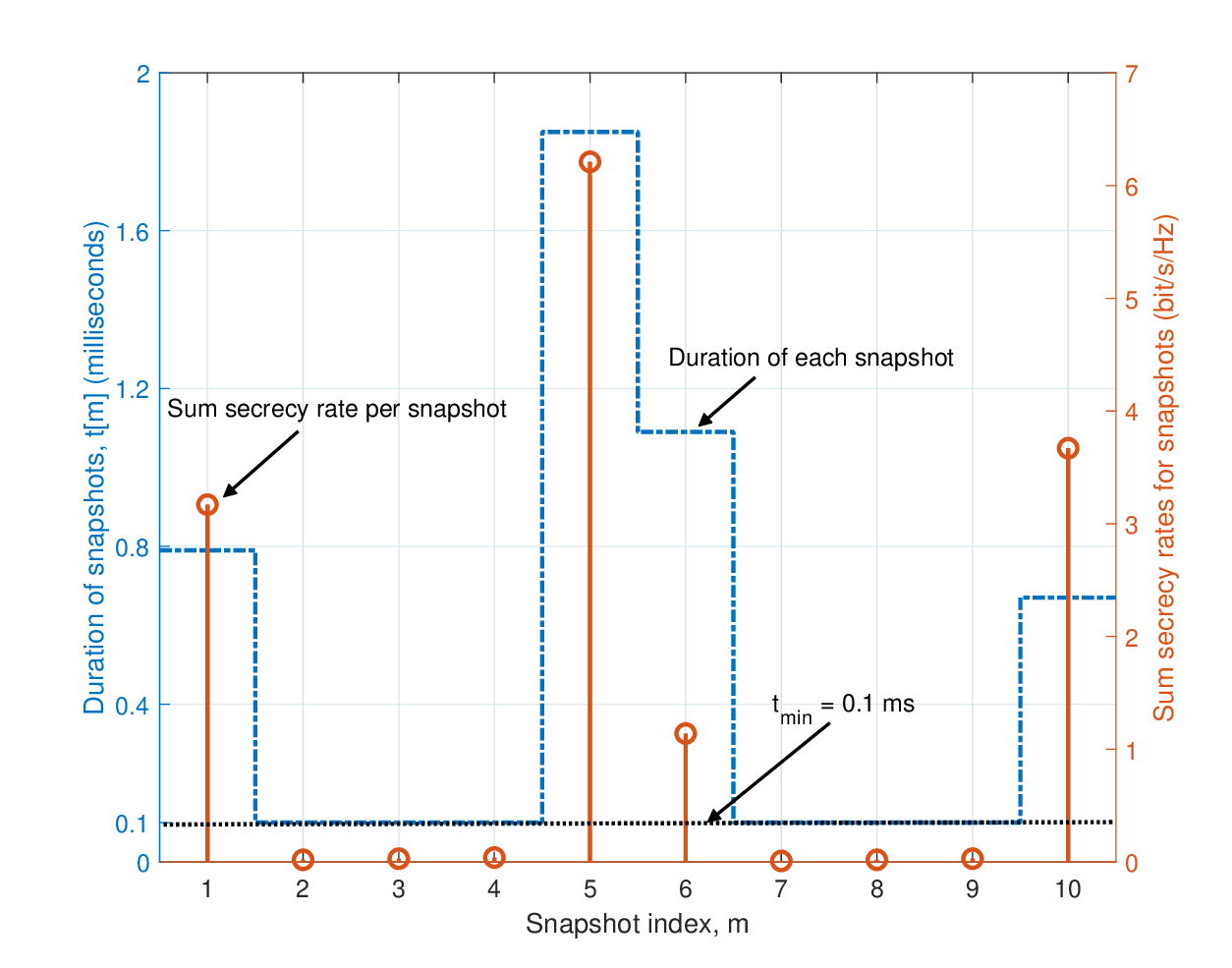}
\caption{Illustration of the durations of the snapshots (ms) and the sum secrecy rates during the snapshots (bits/s/Hz) with $P^{\mathrm{max}}=30$ dBm, $\varsigma_d=0.1$, and $\overline{R}_{\mathrm{req}_k}=0.5$ bits/s/Hz. The minimum required snapshot length is denoted by the dashed black line.}
\label{fig:ISAC_time}
\end{minipage}
\end{figure}
In this subsection, we further investigate the performance of the proposed scheme by focusing on the resource allocation policy for one channel realization of the ISAC system. In Figure \ref{fig:ISAC_polar}, we show the polar plot of the beam patterns of snapshots $1$, $2$, $3$, and $5$ as well as the locations of the legitimate users and the potential eavesdroppers. We observe that the highly-directional beam pattern in each snapshot covers one slice of the sector. As a result, both the legitimate users and any new or known sensing targets are covered by the main lobes of the highly-directional beams, which facilitates ISAC. Moreover, in Figure \ref{fig:ISAC_time}, we show the durations of the snapshots and the sum secrecy rates for the individual snapshots for the considered channel realization of the ISAC system. Both the duration and the sum secrecy rate fluctuate in a similar manner across the snapshots. In fact, once the average achievable rate requirements of all legitimate users are satisfied, the DFRC BS allocates the remaining scanning time to the snapshot that contributes most to the sum secrecy rate, i.e., snapshot $5$ for the considered example. In particular, in snapshot $5$, the DFRC BS jointly serves three legitimate users while the two potential eavesdroppers are far away from these legitimate users, cf. Figure \ref{fig:ISAC_polar}. Hence, it is desirable for the DFRC BS to extend the duration of snapshot $5$, since this simultaneously increases the average achievable rates of the three legitimate users without causing severe information leakage to the eavesdroppers. Moreover, we can observe that for the given parameter setting, the sum secrecy rates of all $10$ snapshots are non-negative. In fact, for the considered problem formulation, negative $R^{\mathrm{sec}}_k[m]$ do not occur in any snapshot of the proposed scheme, see \textit{Remark 1}. Besides, the duration of snapshots $2$, $3$, $4$, $7$, $8$, and $9$ is equal to the minimum snapshot duration $t_{\mathrm{min}}$ required for target sensing.
\par
\begin{figure}[t]
\begin{minipage}[b]{0.47\linewidth}
\hspace*{-5mm}\includegraphics[width=3.2in]{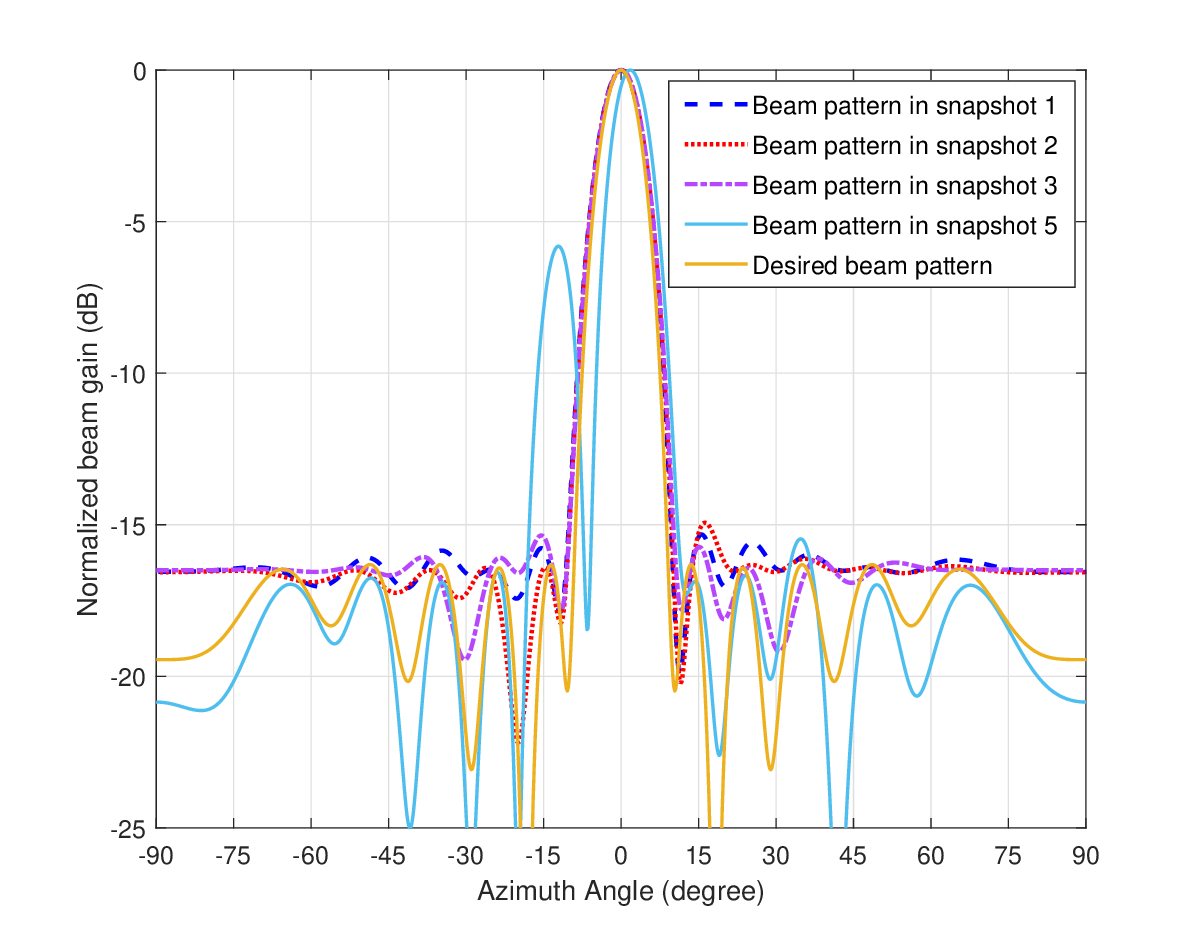}
\caption{Illustration of the normalized beam gain of the desired beam pattern and actual beam patterns in snapshots $1$, $2$, $3$, and $5$.}
\label{fig:ISAC_beampattern}
\end{minipage}\hspace*{8mm}
\begin{minipage}[b]{0.47\linewidth}
\hspace*{-5mm}\includegraphics[width=3.2in]{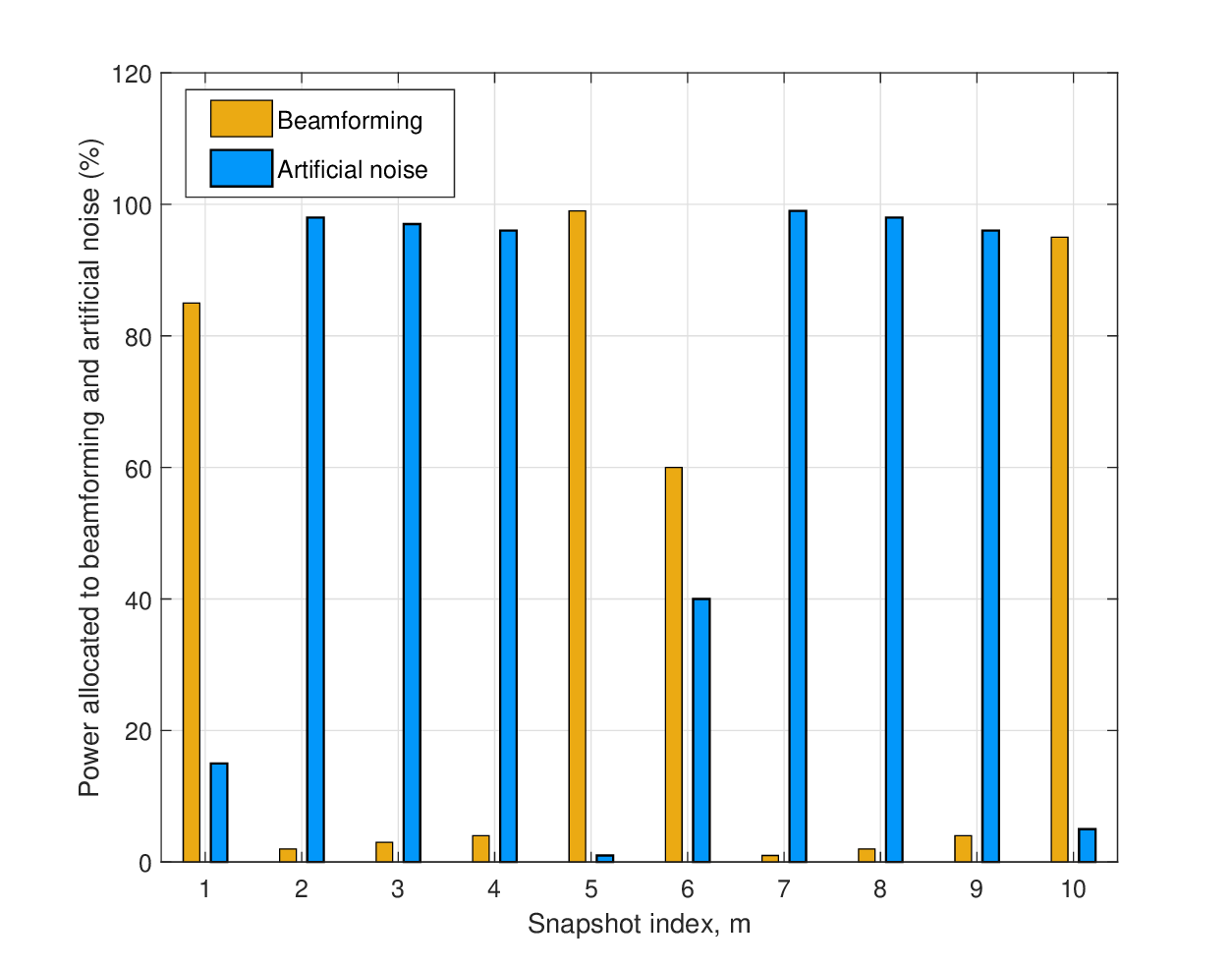}
\caption{The optimized power allocation policy of DFRC BS in each snapshot of a scanning period.}
\label{fig:ISAC_proportion}
\end{minipage}
\end{figure}
In Figure \ref{fig:ISAC_beampattern}, we illustrate the desired beam pattern and the actual beam patterns of snapshots $1$, $2$, $3$, and $5$. The desired beam pattern is obtained by solving an SDP problem similar to \cite[Eq. (13)]{9199556}. To facilitate a comparison, we shift the actual beam patterns such that they are centered at $0$. As can be observed from Figure \ref{fig:ISAC_beampattern}, the actual beam patterns of the first three snapshots accurately approximate the desired beam pattern. In fact, in the presence of angle uncertainties for the potential eavesdroppers, the scheme employing the single snapshot multi-stage design approach would broaden the main lobe of the synthesized beam pattern to impair the channel of potential eavesdroppers for all possible angle uncertainties, cf. \cite{9199556}. In contrast, thanks to the sequential scanning mechanism of the proposed scheme, the DFRC BS is able to maintain a highly-directional synthesized beam, which is favorable for high-quality sensing. As for the beam pattern of snapshot $5$, apart from the main lobe that closely approximates the desired beam pattern, there is also a large sidelobe. Recall that in Figure \ref{fig:ISAC_time}, snapshot $5$ has the longest duration and contributes most to the sum secrecy rate. In fact, the sum secrecy rate mainly benefits from the two legitimate users covered by the main lobe of the beam pattern in snapshot $5$. Hence, for the sake of maximizing the system sum secrecy rate, rather than extending the duration of another snapshot dedicated to serving the legitimate user denoted by the arrow in Figure \ref{fig:ISAC_polar}, the DFRC BS prefers to jointly serve three legitimate users in snapshot $5$. Yet, restricted by the maximum transmit power allowance $P^{\mathrm{max}}$ and the normalized beam pattern difference tolerance factor $\delta_d$, the DFRC BS cannot increase the transmit power and broaden the main lobe of the synthesized beam to cover all three legitimate users. Instead, the DFRC BS exploits the available DoFs to synthesize a suitable sidelobe to satisfy the average achievable rate requirement of the legitimate user denoted by the arrow in Figure \ref{fig:ISAC_polar}. Furthermore, Figure \ref{fig:ISAC_proportion} illustrates how much power the DFRC BS allocates in each snapshot to beamforming and AN. Thanks to the proposed optimization framework, the DFRC BS can flexibly synthesize the desired highly-directional sensing beams by jointly optimizing the beamforming vectors and the AN (e.g., in snapshot 6) or by transmitting only AN (e.g., in snapshot 2) or by using only information beamforming (e.g., in snapshot 5). In particular, for slices containing no users, the DFRC BS approximates the desired sensing beam pattern by using only AN, as this avoids information leakage to the known as well as yet to be discovered eavesdroppers. On the other hand, if there is at least one legitimate user in a slice, the DFRC BS embeds the corresponding information signal in the sensing beam to satisfy the average achievable rate constraint of that legitimate user and to boost the system sum secrecy rate.
\par
\begin{figure}[t]
\centering
\includegraphics[width=3.2in]{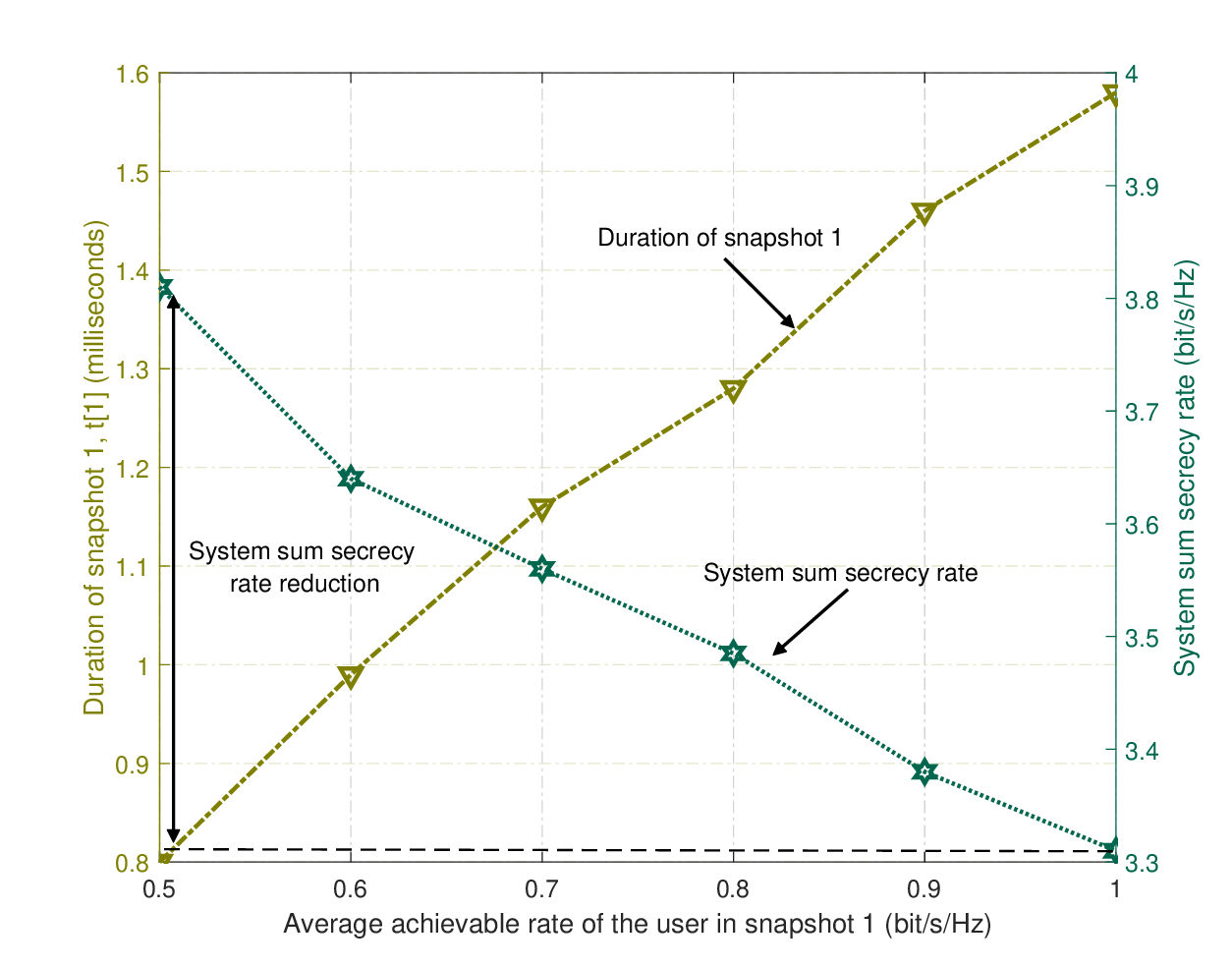}
\caption{Duration of snapshot $1$ (ms) and system sum secrecy rate (bits/s/Hz) versus the average achievable rate of the legitimate user in snapshot $1$ with $P^{\mathrm{max}}=30$ dBm and $\varsigma_d=0.1$.}\label{fig:ISAC_time_snapshot1}
\end{figure}
In Figure \ref{fig:ISAC_time_snapshot1}, we investigate the duration of snapshot $1$ versus the average achievable rate of the legitimate user in snapshot $1$ for one channel realization of the ISAC system. In particular, we gradually increase the average achievable rate of the legitimate user in the area covered by snapshot $1$, i.e., $\overline{R}_{\mathrm{req}_1}$, from $0.5$ bits/s/Hz to $1$ bits/s/Hz and keep $\overline{R}_{\mathrm{req}_k}=0.5$ bits/s/Hz, $k\in\mathcal{K}\setminus\left\{1\right\}$, for the other legitimate users. As expected, the duration of snapshot $1$ increases with $\overline{R}_{\mathrm{req}_1}$. This is because the DFRC BS has to extend the duration of snapshot $1$ to satisfy the more stringent QoS requirements of the legitimate user. On the other hand, we also show the system sum secrecy rate (bits/s/Hz) of the entire scanning period versus $\overline{R}_{\mathrm{req}_1}$. In particular, as $\overline{R}_{\mathrm{req}_1}$ increases from $0.5$ bits/s/Hz to $1$ bits/s/Hz, there is an obvious system sum secrecy rate reduction. This is due to the fact that as the time allocated to snapshot $1$ increases, less time is available for the other snapshots, and particularly for snapshot $5$, which contributes most to the sum secrecy rate, leading to a performance degradation.
\vspace*{0mm}
\section{Conclusions}
\label{ISAC_conclusion}
\vspace*{0mm}
In this paper, we have studied the robust and secure resource allocation algorithm design in ISAC systems over a sequence of snapshots. In particular, the considered DFRC BS periodically scans a service sector by taking a number of snapshots and employs highly-directional beams in each snapshot to facilitate simultaneous sensing and secure communication. To fully unleash the potential of ISAC systems, we proposed a novel design concept based on variable-length snapshots, and developed an optimization framework for the available resources of ISAC systems over a series of snapshots. The duration of each snapshot, the beamforming vector, and the covariance matrix of the AN are jointly optimized to maximize the sum secrecy rate over the scanning period. The CSI of the potential eavesdroppers in the current scanning period can be obtained from the echoes in the previous scanning period. We took into account the unavoidable CSI uncertainty for the legitimate user and potential eavesdropper channels and formulated the resource allocation algorithm design as a non-convex optimization problem. To facilitate a robust resource allocation algorithm design, we adopted a bounded uncertainty model for the distance estimation error and established a tractable bound to capture the joint angle and multi-path fading uncertainty of the eavesdroppers. Subsequently, a computationally efficient BCD-based algorithm with guaranteed convergence was developed by capitalizing on IA and SDR. Our simulation results confirmed that the proposed scheme achieves a significantly higher physical layer security performance compared to three baseline schemes. Moreover, our results showed that the CSI uncertainty of both the legitimate users and the potential eavesdroppers should be carefully considered for robust ISAC system design. Furthermore, our results revealed that compared to the widely adopted single snapshot optimization framework, our variable-length snapshot optimization framework not only facilitates the offline design of highly-directional beam patterns that are favorable for high-quality sensing, but also allows us to prioritize communication and sensing depending on the application scenario.
\par
The proposed optimization framework can be extended to a more general framework by also including the possibility of target tracking. The resulting resource allocation problem constitutes an interesting topic for future work.
\vspace*{0mm}
\bibliographystyle{IEEEtran}
\bibliography{Reference_List}

\begin{thebibliography}{10}
\providecommand{\url}[1]{#1}
\csname url@samestyle\endcsname
\providecommand{\newblock}{\relax}
\providecommand{\bibinfo}[2]{#2}
\providecommand{\BIBentrySTDinterwordspacing}{\spaceskip=0pt\relax}
\providecommand{\BIBentryALTinterwordstretchfactor}{4}
\providecommand{\BIBentryALTinterwordspacing}{\spaceskip=\fontdimen2\font plus
\BIBentryALTinterwordstretchfactor\fontdimen3\font minus
  \fontdimen4\font\relax}
\providecommand{\BIBforeignlanguage}[2]{{%
\expandafter\ifx\csname l@#1\endcsname\relax
\typeout{** WARNING: IEEEtran.bst: No hyphenation pattern has been}%
\typeout{** loaded for the language `#1'. Using the pattern for}%
\typeout{** the default language instead.}%
\else
\language=\csname l@#1\endcsname
\fi
#2}}
\providecommand{\BIBdecl}{\relax}
\BIBdecl

\bibitem{wong2017key}
V.~W.~S. {Wong}, R.~{Schober}, D.~W.~K. {Ng}, and L.~{Wang}, \emph{Key
  Technologies for 5G Wireless Systems}.\hskip 1em plus 0.5em minus 0.4em\relax
  Cambridge University Press, 2017.

\bibitem{7289385}
A.~{Khawar}, A.~{Abdelhadi}, and T.~C. {Clancy}, ``Coexistence analysis between
  radar and cellular system in {L}o{S} channel,'' \emph{IEEE Antennas and
  Wirel. Propag. Lett.}, vol.~15, pp. 972--975, Oct. 2016.

\bibitem{8999605}
F.~{Liu}, C.~{Masouros}, A.~P. {Petropulu}, H.~{Griffiths}, and L.~{Hanzo},
  ``Joint radar and communication design: Applications, state-of-the-art, and
  the road ahead,'' \emph{IEEE Trans. Commun.}, vol.~68, no.~6, pp. 3834--3862,
  Jun. 2020.

\bibitem{9540344}
J.~A. {Zhang}, F.~{Liu}, C.~{Masouros}, R.~W. {Heath}, Z.~{Feng}, L.~{Zheng},
  and A.~{Petropulu}, ``An overview of signal processing techniques for joint
  communication and radar sensing,'' \emph{IEEE J. Sel. Topics Signal
  Process.}, vol.~15, no.~6, pp. 1295--1315, Nov. 2021.

\bibitem{9755276}
Z.~{Wei}, F.~{Liu}, C.~{Masouros}, N.~{Su}, and A.~P. {Petropulu}, ``Toward
  multi-functional {6G} wireless networks: Integrating sensing, communication,
  and security,'' \emph{IEEE Commun. Mag.}, vol.~60, no.~4, pp. 65--71, Apr.
  2022.

\bibitem{9557830}
W.~{Yuan}, Z.~{Wei}, S.~{Li}, J.~{Yuan}, and D.~W.~K. {Ng}, ``Integrated
  sensing and communication-assisted orthogonal time frequency space
  transmission for vehicular networks,'' \emph{IEEE J. Sel. Topics Signal
  Process.}, vol.~15, no.~6, pp. 1515--1528, Nov. 2021.

\bibitem{8641142}
L.~{Zhou}, F.~{Liu}, C.~{Tian}, C.~{Masouros}, A.~{Li}, W.~{Jiang}, and
  W.~{Luo}, ``Optimal waveform design for dual-functional {MIMO}
  radar-communication systems,'' in \emph{Proc. IEEE Int. Conf. Commun. China
  (ICCC)}, Beijing, China, Aug. 2018, pp. 661--665.

\bibitem{8386661}
F.~{Liu}, L.~{Zhou}, C.~{Masouros}, A.~{Li}, W.~{Luo}, and A.~{Petropulu},
  ``Toward dual-functional radar-communication systems: Optimal waveform
  design,'' \emph{IEEE Trans. Signal Process.}, vol.~66, no.~16, pp.
  4264--4279, Aug. 2018.

\bibitem{9685478}
H.~{Hua}, J.~{Xu}, and T.~X. {Han}, ``Transmit beamforming optimization for
  integrated sensing and communication,'' in \emph{Proc. IEEE Global Commun.
  Conf. (GLOBECOM)}, Madrid, Spain, Dec. 2021, pp. 1--6.

\bibitem{8454491}
Y.~{Zhou}, H.~{Zhou}, F.~{Zhou}, Y.~{Wu}, and V.~C.~M. {Leung}, ``Resource
  allocation for a wireless powered integrated radar and communication
  system,'' \emph{IEEE Wireless Commun. Lett.}, vol.~8, no.~1, pp. 253--256,
  Feb. 2019.

\bibitem{9737357}
F.~{Liu}, Y.~{Cui}, C.~{Masouros}, J.~{Xu}, T.~X. {Han}, Y.~C. {Eldar}, and
  S.~{Buzzi}, ``Integrated sensing and communications: Towards dual-functional
  wireless networks for {6G} and beyond,'' \emph{IEEE J. Sel. Areas Commun.},
  vol.~40, no.~6, pp. 1728--1767, Jun. 2022.

\bibitem{9729087}
X.~{Mu}, Y.~{Liu}, L.~{Guo}, J.~{Lin}, and L.~{Hanzo}, ``{NOMA}-aided joint
  radar and multicast-unicast communication systems,'' \emph{IEEE J. Sel. Areas
  Commun.}, vol.~40, no.~6, pp. 1978--1992, Jun. 2022.

\bibitem{9013227}
N.~{Su}, F.~{Liu}, and C.~{Masouros}, ``Enhancing the physical layer security
  of dual-functional radar communication systems,'' in \emph{Proc. IEEE Global
  Commun. Conf. (GLOBECOM)}, Waikoloa, HI, USA, Dec. 2019, pp. 1--6.

\bibitem{1183861}
C.~{Baker} and A.~L. {Hume}, ``Netted radar sensing,'' \emph{IEEE Aerosp.
  Electron. Syst. Mag.}, vol.~18, no.~2, pp. 3--6, Feb. 2003.

\bibitem{negi2005secret}
R.~{Negi} \emph{et~al.}, ``Secret communication using artificial noise,'' in
  \emph{Proc. IEEE Veh. Technol. Conf.}, vol.~62, no.~3, Dallas, USA, Sept.
  2005.

\bibitem{yu2019robust}
X.~{Yu}, D.~{Xu}, Y.~{Sun}, D.~W.~K. {Ng}, and R.~{Schober}, ``Robust and
  secure wireless communications via intelligent reflecting surfaces,''
  \emph{IEEE J. Sel. Areas Commun.}, vol.~38, no.~11, pp. 2637--2652, Nov.
  2020.

\bibitem{4516997}
D.~R. {Fuhrmann} and G.~{San Antonio}, ``Transmit beamforming for {MIMO} radar
  systems using signal cross-correlation,'' \emph{IEEE Trans. Aerosp. Electron.
  Syst.}, vol.~44, no.~1, pp. 171--186, Jan. 2008.

\bibitem{9199556}
N.~{Su}, F.~{Liu}, and C.~{Masouros}, ``Secure radar-communication systems with
  malicious targets: Integrating radar, communications and jamming
  functionalities,'' \emph{IEEE Trans. Wireless Commun.}, vol.~20, no.~1, pp.
  83--95, Jan. 2021.

\bibitem{ren2021optimal}
Z.~{Ren}, L.~{Qiu}, and J.~{Xu}, ``Optimal transmit beamforming for secrecy
  integrated sensing and communication,'' \emph{arXiv:2110.12945}, 2021.

\bibitem{9124713}
X.~{Liu}, T.~{Huang}, N.~{Shlezinger}, Y.~{Liu}, J.~{Zhou}, and Y.~C. {Eldar},
  ``Joint transmit beamforming for multiuser {MIMO} communications and {MIMO}
  radar,'' \emph{IEEE Trans. Signal Process.}, vol.~68, pp. 3929--3944, Jun.
  2020.

\bibitem{1399141}
F.~{Robey}, S.~{Coutts}, D.~W.~J. {McHarg}, and K.~{Cuomo}, ``{MIMO} radar
  theory and experimental results,'' in \emph{Proc. IEEE Asilomar Conference on
  Signals, Systems and Computers}, vol.~1, Pacific Grove, USA, 2004, pp.
  300--304.

\bibitem{4276989}
P.~{Stoica} \emph{et~al.}, ``On probing signal design for {MIMO} radar,''
  \emph{IEEE Trans. Signal Process.}, vol.~55, no.~8, pp. 4151--4161, Aug.
  2007.

\bibitem{4753823}
R.~A.~M. {Fens}, M.~{Ruggiano}, and G.~{Leus}, ``Channel characterization using
  radar for transmission of communication signals,'' in \emph{Proc. European
  Conference on Wireless Technology}, Amsterdam, Netherlands, Oct. 2008, pp.
  127--130.

\bibitem{8738892}
Y.~{Luo}, J.~A. {Zhang}, X.~{Huang}, W.~{Ni}, and J.~{Pan}, ``Optimization and
  quantization of multibeam beamforming vector for joint communication and
  radio sensing,'' \emph{IEEE Trans. Commun.}, vol.~67, no.~9, pp. 6468--6482,
  Sept. 2019.

\bibitem{8365918}
M.~{Chiani}, A.~{Giorgetti}, and E.~{Paolini}, ``Sensor radar for object
  tracking,'' \emph{Proc. IEEE}, vol. 106, no.~6, pp. 1022--1041, Jun. 2018.

\bibitem{8046088}
Y.~{Yang}, R.~B. V.~B. {Simorangkir}, X.~{Zhu}, K.~{Esselle}, and Q.~{Xue}, ``A
  novel boresight and conical pattern reconfigurable antenna with the diversity
  of 360° polarization scanning,'' \emph{IEEE Trans. Antennas Propag.},
  vol.~65, no.~11, pp. 5747--5756, Nov. 2017.

\bibitem{1167263}
W.~{Gawronski} and E.~{Craparo}, ``Antenna scanning techniques for estimation
  of spacecraft position,'' \emph{IEEE Antennas Propag. Mag.}, vol.~44, no.~6,
  pp. 38--45, Dec. 2002.

\bibitem{1221849}
S.~{Kobayashi} and T.~{Iguchi}, ``Variable pulse repetition frequency for the
  global precipitation measurement project ({GPM}),'' \emph{IEEE Trans. Geosci.
  Remote Sens.}, vol.~41, no.~7, pp. 1714--1718, Jul. 2003.

\bibitem{9393560}
J.~{Chen}, J.~{Zhang}, Y.~{Jin}, H.~{Yu}, B.~{Liang}, and D.~{Yang},
  ``Real-time processing of spaceborne {SAR} data with nonlinear trajectory
  based on variable {PRF},'' \emph{IEEE Trans. Geosci. Remote Sens.}, vol.~60,
  pp. 1--12, Apr. 2022.

\bibitem{wang2009worst}
J.~Wang and D.~P. Palomar, ``Worst-case robust {MIMO} transmission with
  imperfect channel knowledge,'' \emph{IEEE Trans. Signal Process.}, vol.~57,
  no.~8, pp. 3086--3100, Aug. 2009.

\bibitem{lorenz2005robust}
R.~G. {Lorenz} and S.~P. {Boyd}, ``Robust minimum variance beamforming,''
  \emph{IEEE Trans. Signal Process.}, vol.~53, no.~5, pp. 1684--1696, Apr.
  2005.

\bibitem{pascual2005robust}
A.~{Pascual-Iserte}, D.~P. {Palomar}, A.~I. {Perez-Neira}, and M.~A. {Lagunas},
  ``A robust maximin approach for {MIMO} communications with imperfect channel
  state information based on convex optimization,'' \emph{IEEE Trans. Signal
  Process.}, vol.~54, no.~1, pp. 346--360, Dec. 2005.

\bibitem{8648450}
S.~{Bi}, J.~{Lyu}, Z.~{Ding}, and R.~{Zhang}, ``Engineering radio maps for
  wireless resource management,'' \emph{IEEE Wireless Commun.}, vol.~26, no.~2,
  pp. 133--141, Apr. 2019.

\bibitem{li2013safe}
Q.~Li, W.-K. Ma, and A.~M.-C. So, ``A safe approximation approach to secrecy
  outage design for {MIMO} wiretap channels,'' \emph{IEEE Signal Process.
  Lett.}, vol.~21, no.~1, pp. 118--121, Dec. 2013.

\bibitem{6860253}
J.~{Xu}, L.~{Liu}, and R.~{Zhang}, ``Multiuser {MISO} beamforming for
  simultaneous wireless information and power transfer,'' \emph{IEEE Trans.
  Signal Process.}, vol.~62, no.~18, pp. 4798--4810, Sept. 2014.

\bibitem{6781609}
D.~W.~K. {Ng}, E.~S. {Lo}, and R.~{Schober}, ``Robust beamforming for secure
  communication in systems with wireless information and power transfer,''
  \emph{IEEE Trans. Wireless Commun.}, vol.~13, no.~8, pp. 4599--4615, Apr.
  2014.

\bibitem{ghosh2010fundamentals}
A.~Ghosh, J.~Zhang, J.~G. Andrews, and R.~Muhamed, \emph{Fundamentals of
  LTE}.\hskip 1em plus 0.5em minus 0.4em\relax Pearson Education, 2010.

\bibitem{8288677}
F.~{Liu}, C.~{Masouros}, A.~{Li}, H.~{Sun}, and L.~{Hanzo}, ``{MU-MIMO}
  communications with {MIMO} radar: From co-existence to joint transmission,''
  \emph{IEEE Trans. Wireless Commun.}, vol.~17, no.~4, pp. 2755--2770, Apr.
  2018.

\bibitem{skolnik1980introduction}
M.~I. {Skolnik}, ``Introduction to radar systems,'' \emph{Avenue of the
  Americas, New York}, 1980.

\bibitem{marks1978general}
B.~R. {Marks} and G.~P. {Wright}, ``A general inner approximation algorithm for
  nonconvex mathematical programs,'' \emph{Operations Research}, vol.~26,
  no.~4, pp. 681--683, 1978.

\bibitem{boyd2004convex}
S.~Boyd and L.~Vandenberghe, \emph{Convex Optimization}.\hskip 1em plus 0.5em
  minus 0.4em\relax Cambridge University Press, 2004.

\bibitem{8310563}
K.~{Shen} and W.~{Yu}, ``Fractional programming for communication
  systems—part {II}: Uplink scheduling via matching,'' \emph{IEEE Trans.
  Signal Process.}, vol.~66, no.~10, pp. 2631--2644, May 2018.

\bibitem{8648498}
Y.~{Sun}, D.~{Xu}, D.~W.~K. {Ng}, L.~{Dai}, and R.~{Schober}, ``Optimal
  3{D}-trajectory design and resource allocation for solar-powered {UAV}
  communication systems,'' \emph{IEEE Trans. Commun.}, vol.~67, no.~6, pp.
  4281--4298, Jun. 2019.

\bibitem{yu2021smart}
X.~{Yu}, V.~{Jamali}, D.~{Xu}, D.~W.~K. {Ng}, and R.~{Schober}, ``Smart and
  reconfigurable wireless communications: From {IRS} modeling to algorithm
  design,'' \emph{IEEE Wireless Commun.}, vol.~28, no.~6, pp. 118--125, Dec.
  2021.

\bibitem{8930608}
Q.~{Wu} and R.~{Zhang}, ``Beamforming optimization for wireless network aided
  by intelligent reflecting surface with discrete phase shifts,'' \emph{IEEE
  Trans. Commun.}, vol.~68, no.~3, pp. 1838--1851, Mar. 2020.

\bibitem{9183907}
D.~{Xu}, X.~{Yu}, Y.~{Sun}, D.~W.~K. {Ng}, and R.~{Schober}, ``Resource
  allocation for {IRS}-assisted full-duplex cognitive radio systems,''
  \emph{IEEE Trans. Commun.}, vol.~68, no.~12, pp. 7376--7394, Dec. 2020.

\bibitem{grant2008cvx}
M.~Grant and S.~Boyd, ``{CVX}: Matlab software for disciplined convex
  programming, version 2.1,'' \emph{http://cvxr.com/cvx}, Jan. 2020.

\bibitem{razaviyayn2013unified}
M.~{Razaviyayn}, M.~{Hong}, and Z.~Q. {Luo}, ``A unified convergence analysis
  of block successive minimization methods for nonsmooth optimization,''
  \emph{SIAM Journal on Optimization}, vol.~23, no.~2, pp. 1126--1153, 2013.

\bibitem{arora2009computational}
S.~{Arora} and B.~{Barak}, \emph{Computational Complexity: A Modern
  Approach}.\hskip 1em plus 0.5em minus 0.4em\relax Cambridge, England:
  Cambridge University Press, 2009.

\bibitem{polik2010interior}
I.~P{\'o}lik and T.~Terlaky, ``Interior point methods for nonlinear
  optimization,'' \emph{Nonlinear Optimization}, pp. 215--276, 2010.

\end{thebibliography}
\end{document}